\documentclass[10pt,journal,compsoc]{IEEEtran}
\ifCLASSOPTIONcompsoc
  \usepackage[nocompress]{cite}
\else
  \usepackage{cite}
\fi
\usepackage{multibib}
\newcites{supp}{Appendix References}

\ifCLASSINFOpdf
  \usepackage[pdftex]{graphicx}
  \graphicspath{{figures/}{../figures/}}
  \DeclareGraphicsExtensions{.pdf,.jpeg,.png}
\else
\fi

\usepackage{amsmath,amsfonts}
\usepackage{amssymb}
\usepackage{tabularx,booktabs}
\usepackage{array}

\ifCLASSOPTIONcompsoc
  \usepackage[caption=false,font=footnotesize,labelfont=sf,textfont=sf]{subfig}
\else
  \usepackage[caption=false,font=footnotesize]{subfig}
\fi
\begingroup\expandafter\expandafter\expandafter\endgroup
\expandafter\ifx\csname IncludeInRelease\endcsname\relax
  \usepackage{fixltx2e}
\fi
\usepackage{placeins}
\usepackage{csquotes}
\usepackage{fontawesome}
\usepackage{pifont}
\usepackage{enumitem}

\usepackage{algorithm}
\usepackage{dsfont}
\usepackage[noend]{algpseudocode}

\usepackage[dvipsnames]{xcolor}
\usepackage{hyperref} 

\ifdefined\showrevision

\usepackage{xcolor}
\usepackage[normalem]{ulem}
\usepackage{marginnote}
\usepackage{xspace}

\newcommand{\revref}[2]{%
\marginnote{$R_{#1}C_{#2}$}
}

\newcommand{\revmod}[1]{%
{\color{blue}#1}\xspace%
}

\newcommand{\revnew}[1]{%
{\color{orange}#1}\xspace%
}

\newcommand{\revdel}[1]{%
{\color{darkgray}\sout{#1}}\xspace%
}

\else

\newcommand{\marginnote}[1]{\ignorespaces}
\newcommand{\revref}[2]{\ignorespaces}
\newcommand{\revmod}[1]{#1}
\newcommand{\revnew}[1]{#1}
\newcommand{\revdel}[1]{\ignorespaces}

\fi

\newcolumntype{Y}{>{\centering\arraybackslash}X}
\newcolumntype{C}[1]{>{\centering\let\newline\\\arraybackslash\hspace{0pt}}m{#1}}
\newcommand{\code}{\href{https://github.com/LucasFidon/trustworthy-ai-fetal-brain-segmentation}{here}}
\newcommand{\cmark}{\textcolor{ForestGreen}{\ding{51}}}
\newcommand{\xmark}{\textcolor{red}{\ding{55}}}
\newcommand{\numraters}{eight}  
\newcommand{\numscoring}{four}

\newcommand{\uzl}{University Hospital Leuven}
\newcommand{\uzlshort}{UHL}

\newcommand{\vienna}{Medical University of Vienna}
\newcommand{\viennashort}{MUV}

\newcommand{\kcl}{King's College London}
\newcommand{\kclshort}{KCL}

\newcommand{\uclh}{University College London Hospital}
\newcommand{\uclhshort}{UCLH}

\newcommand{\manchester}{Manchester}
\newcommand{\manchestershort}{MCT}

\newcommand{\belfast}{Belfast}
\newcommand{\belfastshort}{BFT}

\newcommand{\cork}{Cork}
\newcommand{\corkshort}{CRK}

\newcommand{\newcastle}{Newcastle}
\newcommand{\newcastleshort}{NCS}

\newcommand{\liverpool}{Liverpool}
\newcommand{\liverpoolshort}{LVP}

\def\TWAI{\textrm{TWAI}}
\def\AI{\textrm{AI}}
\def\fallback{\textrm{fallback}}
\def\failsafe{\textrm{fail-safe}}
\def\anatomy{\textrm{anatomy}}
\def\intensity{\textrm{intensity}}

\begin{document}
%
\title{A Dempster-Shafer approach to trustworthy AI with application to fetal brain MRI segmentation}
%
%
%
%

\author{
Lucas Fidon,
Michael Aertsen,
Florian Kofler,
Andrea Bink,
Anna L. David,
Thomas Deprest,
Doaa Emam,
Fr\'ed\'eric Guffens,
Andr\'as Jakab,
Gregor Kasprian,
Patric Kienast,
Andrew Melbourne,
Bjoern Menze,
Nada Mufti,
Ivana Pogledic,
Daniela Prayer,
Marlene Stuempflen,
Esther Van Elslander,\\
S\'ebastien Ourselin,
Jan Deprest,
Tom Vercauteren
\IEEEcompsocitemizethanks{
\IEEEcompsocthanksitem 
L. Fidon, A. Melbourne, N. Mufti, S. Ourselin, and T. Vercauteren were with 
the School of Biomedical Engineering \& Imaging Sciences, King's College London, London, UK.
M. Aertsen, T. Deprest, D.Emam, F. Guffens, E. van Elslander, and J. Deprest were with the
Department of Radiology, University Hospitals Leuven, Leuven, Belgium.
F. Kofler and B. Menze were with the
Department of Informatics, Technical University Munich, Munich, Germany.
A. Bink was with the
Department of Neuroradiology and Clinical Neuroscience Center, University Hospital Zurich and University of Zurich, Zurich, Switzerland.
A. L. David was with the
Institute for Women’s Health, University College London, London, UK.
D. Emam was with the
Department of Gynecology and Obstetrics, University Hospitals Tanta, Tanta, Egypt.
A. Jakab was with the
Center for MR Research, University Children’s Hospital Zurich, University of Zurich, Zurich, Switzerland.
G. Kasprian, P. Kienast, I. Pogledic, D. Prayer, and M. Stuempflen were with the
Department of Biomedical Imaging and Image-guided Therapy, Medical University of Vienna, Vienna, Austria.
\protect\\
E-mail: lucas.fidon@kcl.ac.uk
}
}

\IEEEtitleabstractindextext{%
\begin{abstract}
    Deep learning models for medical image segmentation can fail unexpectedly and spectacularly for pathological cases and images acquired at different centers than training images, with labeling errors that violate expert knowledge.
    %
    Such errors undermine the trustworthiness of deep learning models for medical image segmentation.
    Mechanisms for detecting and correcting such failures are essential for safely translating this technology into clinics and are likely to be a requirement of future regulations on artificial intelligence~(AI). 
    In this work, we propose a trustworthy AI theoretical framework and a practical system that can augment any backbone AI system using a fallback method and a fail-safe mechanism based on Dempster-Shafer theory. Our approach relies on an actionable definition of trustworthy AI.
    Our method automatically discards the voxel-level labeling predicted by the backbone AI that violate expert knowledge and relies on a fallback for those voxels.
    We demonstrate the effectiveness of the proposed trustworthy AI approach on the largest reported annotated dataset of fetal MRI consisting of $540$ manually annotated fetal brain 3D T2w MRIs from $13$ centers.
    %
    Our trustworthy AI method improves the robustness of 
    \revmod{four backbone AI models}
    for fetal brain MRIs acquired across various centers and for fetuses with various brain abnormalities.
    Our code is publicly available \code{}.
\end{abstract}

}

\maketitle

\IEEEdisplaynontitleabstractindextext

%
\IEEEpeerreviewmaketitle

\IEEEraisesectionheading{\section{Introduction}\label{sec:introduction}}
%

\begin{figure*}[tp]
    \centering
    \includegraphics[width=0.9\textwidth, trim=0cm 0cm 0cm 4cm]{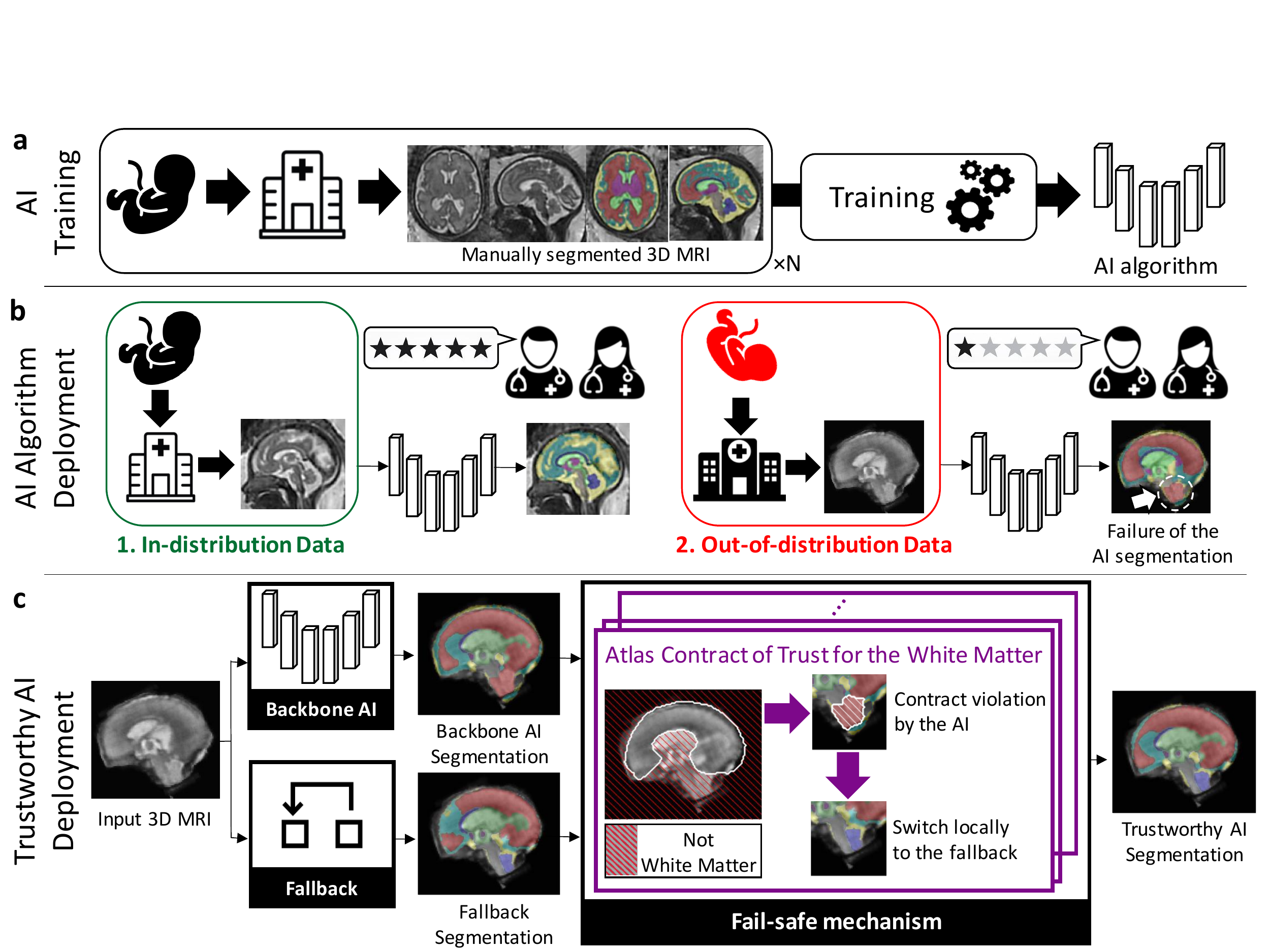}
    \caption{
    \textbf{Schematics of our principled method for trustworthy AI applied to medical image segmentation.}
    \textbf{a.} Deep neural networks for medical image segmentation (AI algorithm) are typically trained on images from a limited number of acquisition centers. 
    This is usually not sufficient to cover all the anatomical variability.
    \textbf{b.} When such a trained AI algorithm is deployed, it will typically give satisfactory accuracy for images acquired with the same protocol as training images and with a health condition represented in the training dataset (left).
    However, an AI algorithm might fail with errors that are not anatomically plausible, for images acquired with a slightly different protocol as training images and/or representing anatomy underrepresented in the training dataset (right).
    \textbf{c.} Schematic of the proposed trustworthy AI algorithm.
    A backbone AI segmentation algorithm is coupled with a \emph{fallback} segmentation algorithm.
    Experts knowledge about the anatomy is modeled in the \emph{fail-safe mechanism} using Dempster-Shafer theory.
    A rich variety of experts knowledge can be modeled as \emph{contracts of trust}, such as, but not only, atlas-based prior and intensity-based prior (not shown here).
    When part of the AI segmentation is found to contradict one of the contracts of trust for a voxel, our trustworthy AI algorithm automatically switches continuously to the fallback segmentation for this voxel.
    }
    \label{fig:overview}
\end{figure*}

\IEEEPARstart{A}{utomatic} segmentation of medical images is needed for personalized medicine and to study anatomical development in healthy populations as well as populations with a pathology.
Artificial Intelligence (AI) algorithms for medical image segmentation can reach super-human accuracy on average~\cite{isensee2021nnu} and yet most radiologists do not trust them~\cite{allen20212020,cabitza2019biases}.
This is partly because, for some cases, AI algorithms fail spectacularly with errors that violate expert knowledge about the segmentation task when the AI was applied across imaging protocol and anatomical pathologies~\cite{allen20212020,fidon2021distributionally,gonzalez2021detecting} (Fig.\ref{fig:overview}b).
This sense of distrust is exacerbated by the current lack of clear
fit-for-purpose
regulatory requirements for AI-based medical image software~\cite{van2021artificial}.

The legal framework for the deployment in clinics of AI tools for medical segmentation is likely to soon become more stringent once the European Union has proposed its Artificial Intelligence Act to regulate AI
and AI trust is at the core of this proposal~\cite{ethics,regulations2021}.
%
%
%
%
%
Guidelines for trustworthy AI claim that AI trustworthiness must precede trust in the deployment of AI systems to avoid miscalibration of the human trust with respect to the trustworthiness of an AI system~\cite{ethics}.
In Psychology, trust of humans in AI can be defined as the belief of the human user that the AI system will satisfy the criteria of a set of contracts of trust.
This contract-based definition of human-AI trust reflects the plurality and the context-dependency of human-AI trust. In particular, the user may trust an AI system for one population or one type of scanner but not trust it for others.
An AI system is trustworthy to a contract of trust if it can maintain this contract
in all situations within the contract scope.
The EU ethics guidelines for trustworthy AI, that upheld the AI Act, advocate that
\enquote{AI systems should have safeguards that enable a \textit{fallback} plan in case of problems}~\cite{ethics}.
We argue that those safeguards should implement a \textit{fail-safe} mechanism in relation with a collection of contracts of trust so as to improve the trustworthiness of the overall system.

In this article, we propose the first trustworthy AI framework with a fail-safe and a fallback
for medical image segmentation.
%
The proposed framework consists of three main components: 
first, a backbone AI algorithm, that can be any AI algorithm for the task at hand,
second, a fallback segmentation algorithm, that 
\revmod{\revref{1}{2}
is more robust than the backbone AI algorithm to out-of-distribution data but potentially less precise,
}
%
and third, a fail-safe method that automatically detects local conflicts between the backbone AI algorithm prediction and the contracts of trust and switches to the fallback algorithm in case of conflicts.
This is illustrated for fetal brain 3D MRI segmentation in Fig.~\ref{fig:overview}.
The proposed principled fail-safe method is based on Dempster-Shafer~(DS) theory.
DS theory allows to model partial information about the task, which is typically the case for expert knowledge.
For example, in human brain anatomy, the cerebellum is known to be located in the lower back part of the brain.
This gives us information only about the segmentation of the cerebellum while a segmentation algorithms will typically compute segmentation for many other tissue types in addition to the cerebellum.
The Dempster's rule of combination of DS theory is an efficient mathematical tool to combine independent sources of information that discards contradictions among the sources.
%
In our framework, the AI-based segmentation and each expert knowledge are treated as independent sources of information and the Dempster's rule of combination is employed to act as the fail-safe.

To demonstrate the applicability of the developed trustworthy AI framework,
we propose one implementation for fetal brain segmentation in MRI.
For the backbone AI model, we used the state-of-the-art deep learning-based segmentation pipeline nnU-Net~\cite{isensee2021nnu}.
%
For the fallback model we used a registration-based segmentation method inspired by the state-of-the-art multi-atlas method GIF~\cite{cardoso2015geodesic}.
%
We also show that our fail-safe formulation is flexible enough to model both spatial location-based 
and intensity-based 
contracts of trust about the regions of interest to be segmented.
%
Spatial location-based fail-safes are implemented using the masks of the regions of interest computed by the registration-based fallback algorithm.
However, in the fail-safe, the masks are interpreted differently.
In this case, the mask of a region R is used only to exclude labeling voxels outside of the mask as belonging to R. 
This is illustrated in Fig.~\ref{fig:overview}c.
Inspired by the margins used for safety in radiation therapy planning to account for spatial registration errors~\cite{niyazi2016estro}, 
we first add spatial margins to the fallback masks before excluding the labels seen as anatomically unlikely according to the dilated fallback masks.
While allowing the masks to overlap, this helps preventing miscoverage of the regions of interest that is the only source of error in this formulation of the fail-safe.

We evaluated the proposed trustworthy AI method on fetal brain segmentation into eight tissue types using 3D T2w MRI.
The segmentation of fetal brain MRI is essential to study normal and abnormal fetal brain development.
In the future, reliable analysis and evaluation of fetal brain structures could also support diagnosis of central nervous system pathology, patient selection for fetal surgery, evaluation and prediction of outcome, hence also parental counselling. 
In particular, fetal brain 3D T2w MRI segmentation presents multiple challenges for trustworthy AI~\cite{fidon2021distributionally}.
There are variations in T2w MRI protocols used across clinical centers and there is a spectacular variation of the fetal brain anatomy across gestational ages and across normal and abnormal fetal brain anatomy.

\section{Related Works}

\subsection{Information fusion for medical image segmentation}

Information fusion methods based on probability theory have been proposed to combine different segmentations~\cite{warfield2004simultaneous,welinder2010multidimensional}.
The Simulataneous Truth And Performance Level Estimation (STAPLE) algorithm weighs each segmentation by estimating the sensitivity and specificity of each segmentations~\cite{warfield2004simultaneous}.
In particular, these methods define only image-wise weights to combine the segmentations.
Fusion methods with weights varying spatially have been proposed for the special case of atlas-based algorithms~\cite{cardoso2015geodesic}, but not in general as in our method.
In the context of deep learning-based segmentation methods, simple averaging is used in state-of-the-art pipelines~\cite{isensee2021nnu}.
Perhaps more importantly, fusion methods based on probability theory only cannot model imprecise or partial prior expert-knowledge~\cite{warfield2004simultaneous}.
In contrast, the use of Dempster-Shafer (DS) theory in our method allows us a larger diversity of prior knowledge that is typically robust but imprecise.
We show that our approach based on DS can model prior given by either atlases or voxel intensity prior distributions in the case of fetal brain segmentation and more priors could be modeled as well in other segmentation tasks.

\subsection{Dempster-Shafer for medical image segmentation}

Only a few papers have proposed to use DS theory in the context of information fusion for medical image segmentation~\cite{bloch1996some,capelle2004evidential,ghasemi2013novel,lelandais2014dealing,liu2015new}.
DS has been used to combine different image modalities~\cite{bloch1996some}, neighbouring voxels~\cite{liu2015new}, or both~\cite{capelle2004evidential,ghasemi2013novel} for brain MRI segmentation.

In contrast, in this work DS is used to combine two arbitrary probabilistic segmentation algorithms with prior information about the segmentation tasks.
In this case, we show that Dempster's rule of combination allows to detect segmentation failures of the first segmentation algorithms and switch to the second locally at the voxel level.

\revnew{
\subsection{Domain generalization}\revref{2}{1}
Domain generalization (DG) aims at improving the out-of-distribution (OOD) generalization of AI algorithms without using OOD images during training~\cite{zhou2022domain}.
DG methods include the use of various data augmentations~\cite{zhang2020generalizing,isensee2021nnu}, self-supervised learning pre-training~\cite{tang2022self}, and new training methods~\cite{fidon2020distributionally}.
Closer to this work are DG methods based on the integration of anatomical prior~\cite{kushibar2018automated,liu2022single}.
In those methods, atlas-based probabilities are fused using concatenation inside the deep neural network.

DG can be used to improve generalizability but it does not offer any guarantees in terms of trustworthiness.
As such, it is a complementary approach to the one we propose.
In our experiments, our approach is compared and combined with three DG methods as backbone AI leading to improved robustness.
}

\section{Methods}


\subsection{Background on Human-AI trust}\label{sec:humanAItrust}
%
%
\textbf{Artificial intelligence} is defined as any automation perceived by the individual using it as having an intent~\cite{jacovi2021formalizing}.

Human-AI trust is multi-dimensional.
For example, the user can trust an AI medical image segmentation algorithm for a given tissue type, for images coming from a given type of scanner, or for a given population and not another.
This observation that trust has several facets and is context-dependent~\cite{hoffman2017taxonomy} has motivated the introduction of \emph{contractual trust}~\cite{jacovi2021formalizing}.
A \textbf{contract of trust} is an attribute of the AI algorithm which, if not fulfilled, causes a risk in using the AI algorithm.
A contract of trust is not necessarily related to the accuracy of the AI algorithm for the task at hand.

For the automatic segmentation of the heart on chest CT images, one contract of trust could be:
"The heart labels are always on the left side of the body."
%
%
The AI algorithm can fulfil this contract and yet compute an inaccurate segmentation of the heart.
This contract can also be restricted to CT images of sufficient quality.
%
%
Context can be added to the contract and several contracts can be derived from the contract above for different contexts.
%
In the previous example, context is implied in that
this contract does not apply to individuals with dextrocardia.

An AI algorithm is defined as \textbf{trustworthy} with respect to a contract of trust
if it provides guarantees that it will abide by the obligations of said contract~\cite{jacovi2021formalizing}.

The requirements proposed in the EU guidelines for trustworthy AI~\cite{ethics} are examples of contracts of trust.
One important requirement of trustworthy AI is the technical robustness and safety~\cite{ethics} which is the focus of our work.
The EU guidelines propose to achieve trustworthiness in practice using a \emph{fallback plan}.
However, no technical means of implementing such plan have been provided or published. 
Problems with the backbone AI algorithm should be detected using a \emph{fail-safe} algorithm and a \emph{fallback} algorithm should be available.
In the remaining of this section we will present a theoretical framework for the implementation of a trustworthy AI system leveraging Dempster-Shafer theory to implement a failsafe and a fallback plan.
And we will show how our framework can be used to maintain several concrete contracts of trust for fetal brain MRI segmentation.


\subsection{Background on Dempster-Shafer theory}\label{sec:DS}
%
%
In Dempster-Shafer theory~\cite{shafer1976mathematical}, \textit{basic probability assignments} are a generalization of probabilities that allow to model partial and imprecise information and to combine different sources of information using Dempster's rule.

Let $\mathbf{C}$ be the set of all classes and $2^{\mathbf{C}}$ the set of all subsets of $\mathbf{C}$.
A \textbf{basic probability assignment (BPA)} on $\mathbf{C}$ is a function $m: 2^{\mathbf{C}} \mapsto [0,\,1]$ that satisfies
\begin{equation}
    m(\emptyset) = 0 
    \quad \textup{and} \quad
    \sum_{A \subset \mathbf{C}} m(A) = 1
\label{eq:bpa}
\end{equation}
Probabilities on $\mathbf{C}$ are functions $p: \mathbf{C} \mapsto [0,1]$ that satisfy $\sum_{c\in \mathbf{C}}p(c) = 1$.
Probabilities are equivalent to the BPAs that assign non-zeros weights only to the singletons, i.e. the sets $A=\{c\}$ for $c \in \mathbf{C}$.
Given a probability $p$ the BPA $m^{(p)}$ associated with $p$ is defined as: 
$\forall c \in \mathbf{C},\, m^{(p)}(\{c\})=p(c)$ 
and $\forall A \subset \mathbf{C}$ with $|A| \neq 1$, $m^{(p)}(A)=0$.
Basic probability assignments are therefore more general than probabilities.

For $A \subset \mathbf{C}$, $m(A)$ is the probability that our knowledge about the true label is exactly and only: "the true class is one of the classes in $A$".
In particular, it does not imply that $m(B)>0$ for any set $B$ such that $B \subsetneq A$ or $A \subsetneq B$.
This is in contrast to probabilities that can weight only the individual classes.
BPAs allow to represent more precisely than probabilities what we know (and don't know) about the true class of a voxel.
For example, the extreme case where we know nothing about the true class can be represented by the BPA $m$ such that $m(\mathbf{C})=1$.
The best one can do to try representing this case with probabilities is to define a probability $p$ such that $\forall c \in \mathbf{C},\, p(c)= \frac{1}{|\mathbf{C}|}$.
However, this choice of $p$ corresponds to the knowledge that the class distribution is uniformly random which is different from knowing nothing about the true class.

%
Finally, two BPAs on $\mathbf{C}$, $m_1$ and $m_2$, are said to be \textbf{completely contradictory} if 
\begin{equation}
\label{eq:contradiction}
    \sum_{E,F\subset \mathbf{C}|E \cap F = \emptyset} m_1(E)m_2(F) = 1
\end{equation}
Using \eqref{eq:bpa}, $m_1$ and $m_2$ are completely contradictory if and only if one cannot form a pair of overlapping sets of classes $(A,B)$ such that 
$m_1$ commits some belief to $A$, i.e. $m_1(A) > 0$, 
and $m_2$ commits some belief to $B$, i.e. $m_2(B) > 0$.

\subsubsection{Dempster's rule of combination}
Dempster's rule of combination allows to combine any pair $(m_1, m_2)$ of BPAs on $\mathbf{C}$ that are not completely contradictory using the formula, 
$\forall A \subset \mathbf{C},$
\begin{equation}
    (m_1 \oplus m_2)(A) =
    \left\{
    \begin{array}{cc}
        \frac{\sum_{E,F\subset \mathbf{C}|E \cap F = A} m_1(E)m_2(F)}{1 - \sum_{E,F\subset \mathbf{C}|E \cap F = \emptyset} m_1(E)m_2(F)} & \textup{if}\,\, A \neq \emptyset\\
        0 & \textup{if}\,\, A = \emptyset
    \end{array}
    \right.
    \label{eq:ds}
\end{equation}
%
It is worth noting that $m_1 \oplus m_2$ is also a BPA on $\mathbf{C}$.
In addition, the relation $\oplus$ is symmetrical and associative.

One particular case that will be useful for our method is the combination of a probability $p$ on $\mathbf{C}$ with a generic BPA $m$ on $\mathbf{C}$ using Dempster's rule of combination.

Since $p$ is a probability, for $A\subset\mathbf{C}$, $p(A)$ can be non-zeros only if $A$ is a singleton, i.e. if it exists a class $c \in \mathbf{C}$ such that $A=\{c\}$.
For simplicity, we will therefore use the abusive notation when considering $p$ as a BPA:
$p(c):=p(A)=p(\{c\})$.
The relation of complete contradiction \eqref{eq:contradiction} between $p$ and $m$ can be simplified
\begin{equation}
\label{eq:contradiction_proba}
    \sum_{E,F\subset \mathbf{C}|E \cap F = \emptyset} p(E)m(F)
    = \sum_{c \in \mathbf{C}} \sum_{F \subset \left(\mathbf{C}\setminus \{c\}\right)} p(c)m(F)
    = 1
\end{equation}

Similarly, if $p$ and $m$ are not completely contradictory, the Dempster's rule between $p$ and $m$ can be simplified.
Let $A \subset \mathbf{C}$, $A\neq \emptyset$, using \eqref{eq:ds} we have
%
\begin{equation}
    (p \oplus m)(A) = \frac{\sum_{c \in \mathbf{C},F\subset \mathbf{C}|\{c\} \cap F = A} p(c)m(F)}{1 - \sum_{c \in \mathbf{C}} \sum_{F \subset \left(\mathbf{C}\setminus \{c\}\right)} p(c)m(F)}
\end{equation}

We remark that $p\oplus m$ is also a probability on $\mathbf{C}$.
Indeed, let $c\in \mathbf{C}$, it can exist $F \subset \mathbf{C}$ such that $\{c\} \cap F = A$ only if $A$ is the singleton $A=\{c\}$ (we have assumed $A \neq \emptyset$).
Therefore if $A$ is not a singleton, i.e. $|A|>1$, the sum on the numerator is empty and equal to $0$.
As a result, $\forall c \in \mathbf{C}$,
\begin{equation}
    (p \oplus m)(c) = \frac{p(c)\sum_{F\subset \mathbf{C}|c \in F} m(F)}{1 - \sum_{c' \in \mathbf{C}} \sum_{F \subset \left(\mathbf{C}\setminus \{c'\}\right)} p(c')m(F)}
    \label{eq:drc_proba}
\end{equation}

\subsection{A Dempster-Shafer approach to Trustworthy AI}\label{sec:TWAI}
%
%
Our trustworthy AI segmentation method consists of three main components:
1) a backbone AI segmentation algorithm; 2) a fallback segmentation algorithm; and 3) and 
a fail-safe method that detects area of conflict between the AI algorithm segmentation and the contracts of trust and switches to the fallback algorithm for those regions.
An illustration is given in Fig.~\ref{fig:overview}.

The AI segmentation algorithm is a high-accuracy segmentor that can be, for example, a state-of-the-art convolutional neural network.
%
The fallback segmentation algorithm is a segmentor that might achieve lower accuracy than the AI, but is superior to the AI for other desirable properties such as robustness.
It is worth noting that AI and fallback segmentors are interchangeable in theory, and that either of them could consist of manual or semi-automatic segmentation methods.
The AI and fallback segmentation algorithms take as input an image to be segmented and compute for each voxel of the image a probabilities vector with one probability for each class to be segmented.

The fail-safe mechanism aims at detecting erroneous predictions of the AI segmentation algorithm that contradict one of the contracts of trust.
The contracts of trust embed domain knowledge such as "there can't be white matter in this part of the brain" or "hyperintense voxels on T2 fetal brain MRI are always cerebrospinal fluid".
%
%
%
Most contract will not enforce a specific segmentation but rather impose that the automatic segmentation meets certain constraints.
%
%
In the context of image segmentation, contract of trusts can only reduces the set of possible classes and reweights the class probabilities of the segmentation of a pixel or voxel.
To implement the fail-safe mechanism, we propose to use a basic probability assignment (BPA) that acts on the backbone AI and the fallback class probabilities using Dempster's rule of combination~\eqref{eq:drc_proba}.
In addition, we assume that the fallback class probabilities never completely contradict the BPA representing the contracts of trust.
As a result, Dempster's rule of combination can be used to switch automatically between the backbone AI algorithm and the fallback algorithm when the AI class probabilities completely contradict the BPA.
%
%
Formally, the trustworthy segmentation prediction is defined for an input image $I$ and for all voxel position $\textbf{x}$ as
\begin{equation}
\label{eq:trustworhtyAI}
    p^{\TWAI}_{I, \textbf{x}} = 
    \left(
    (1 - \epsilon) p^{\AI}_{I, \textbf{x}} + \epsilon p^{\fallback}_{I, \textbf{x}}
    \right)
    \oplus m^{\failsafe}_{I, \textbf{x}}
\end{equation}
where $\oplus$ is the Dempster's combination rule \eqref{eq:ds},
$p^{\AI}_{I, \textbf{x}}$ is the class probability prediction of the AI segmentation algorithm for voxel $\textbf{x}$ of image $I$,
$p^{\fallback}_{I, \textbf{x}}$ is the class probability prediction of the fallback segmentation for voxel $\textbf{x}$ of image $I$,
and $m^{\failsafe}_{I, \textbf{x}}$ is the BPA of the fail-safe mechanism for voxel $\textbf{x}$ of image $I$.
The parameter $\epsilon$ is a constant in $]0,1]$.
A toy example is given in Appendix \ref{sec:toy-example}

\subsubsection{Fail-safe mechanism}\label{sec:fail-safe}
In our framework, we assume that the fallback segmentation algorithm always produces segmentation probabilities that do not contradict entirely the BPA of the contracts of trust.
A trivial example of such fallback, is the uniform segmentation algorithm that assigns an equal probability to all the classes to be segmented and for all voxels.
In contrast, we do not make such compatibility assumption for the AI segmentation algorithm.
Not only does this make our approach applicable with any AI segmentation algorithm, but our method also relies on the incompatibility between the AI segmentation algorithm prediction and the contracts of trust to detect failure of the AI segmentation algorithm and to switch to the fallback segmentation algorithm.
Formally, however small the weight $\epsilon$ given to the fallback is, as long as $\epsilon>0$, when $p^{\AI}_{I, \textbf{x}}$ is completely contradictory with $m^{\failsafe}_{I, \textbf{x}}$, we obtain that $p^{\TWAI}_{I, \textbf{x}}$ depends only on $p^{\fallback}_{I, \textbf{x}}$ and not on $p^{\AI}_{I, \textbf{x}}$.
On the contrary, when $p^{\AI}_{I, \textbf{x}}$ is not completely contradictory with $m^{\failsafe}_{I, \textbf{x}}$, we obtain that $p^{\TWAI}_{I, \textbf{x}}$ depends mainly on $p^{\AI}_{I, \textbf{x}}$ and the contribution of $p^{\fallback}_{I, \textbf{x}}$ is negligible for $\epsilon$ small enough.
%
%
Here, we consider the case in which the AI algorithm predicted probability $p^{\AI}_{I,\textbf{x}}$ is completely contradictory with $m^{\failsafe}_{I,\textbf{x}}$ for a voxel $\textbf{x}$.
Using \eqref{eq:contradiction_proba}
\begin{equation*}
\left\{
    \begin{aligned}
        &\sum_{c'\in \mathbf{C}}
        \left(
        \sum_{\mathbf{C}'\subset (\mathbf{C}\setminus \{c'\})}
        p^{\AI}_{I, \textbf{x}}(c')\,m^{\failsafe}_{I,\textbf{x}}\left(\mathbf{C}'\right)
        \right)
        = 1\\
        & \forall c' \in \mathbf{C},\,\forall \mathbf{C}'\subset \mathbf{C}\,|\, c' \in \mathbf{C}',\quad
        p^{\AI}_{I, \textbf{x}}(c')\,m^{\failsafe}_{I,\textbf{x}}\left(\mathbf{C}'\right) = 0
    \end{aligned}
\right.
\end{equation*}
Using Dempster's rule of combination \eqref{eq:drc_proba} we obtain, 
\begin{equation}
\forall c \in \mathbf{C},\quad
    p^{\TWAI}_{I, \textbf{x}}(c) =
        \left(
        p^{\fallback}_{I, \textbf{x}} \oplus m^{\failsafe}_{I,\textbf{x}}
        \right)(c)
\end{equation}
%
However small $\epsilon>0$ can be, the trustworthy AI prediction for voxel $\textbf{x}$ does not depend anymore on the AI algorithm probability but only on the fallback algorithm probability.
In other words, we have switched totally from the backbone AI algorithm to the fallback algorithm.

\subsubsection{General case with multiple contracts of trust}\label{sec:multi_contracts}
In general, $m^{\failsafe}_{I, \textbf{x}}$ is a sum of contracts of trust BPAs that are not completely contradictory and can be written as
\begin{equation}
    m^{\failsafe}_{I, \textbf{x}} = \bigoplus_{k=1}^K m^{(k)}_{I, \textbf{x}}
\end{equation}
where each $m^{(k)}_{I, \textbf{x}}$ is a basic probability assignment (BPA), $K$ is the number of BPAs, and $\bigoplus_{k=1}^K$ is the Dempster's rule of combination \eqref{eq:ds} of $K$ BPAs computed in any order.
%
%
The $m^{(k)}_{I, \textbf{x}}$ represent the contracts of trust in our framework.

Specifically, for medical image segmentation we propose the following trustworthy AI model:
\begin{equation}
    \label{eq:trustworhtyAI-fetal}
    p^{\TWAI}_{I, \textbf{x}} = 
    \left((1 - \epsilon) p^{\AI}_{I, \textbf{x}} + \epsilon p^{\fallback}_{I, \textbf{x}}\right)
    \oplus m^{\anatomy}_{I, \textbf{x}}
    \oplus m^{\intensity}_{I, \textbf{x}}
\end{equation}
where
$m^{\anatomy}_{I, \textbf{x}}$ is the anatomical contract of trust BPA for voxel $\textbf{x}$ of image $I$,
and $m^{\intensity}_{I, \textbf{x}}$ is the intensity contract of trust BPA for voxel $\textbf{x}$ of image $I$.
The definitions of $m^{\anatomy}_{I, \textbf{x}}$ and $m^{\intensity}_{I, \textbf{x}}$ will be derived in sections \ref{sec:anatomical_contract} and \ref{sec:intensity_contract}.

\subsubsection{Dempster-Shafer anatomical contracts of trust}\label{sec:anatomical_contract}
In this section, we describe our proposed anatomical prior basic probability assignment (BPA) $m^{\anatomy}$ that is used in our trustworthy AI method \eqref{eq:trustworhtyAI-fetal}.

Our anatomical prior is computed using the segmentations computed using a multi-atlas segmentation algorithm~\cite{cardoso2015geodesic}.
Atlas-based segmentation algorithms are anatomically-constrained due to the spatial smoothness that is imposed to the spatial transformation used to compute the segmentation.
In practice, this is achieved thanks to the parameterization of the spatial transformation and the regularization loss in the registration optimization problem~\cite{cardoso2015geodesic,modat2010fast}.
Therefore, if implemented correctly, atlas-based automatic segmentations can inherit from the anatomical prior represented by segmentation atlases.

In terms of contract of trust, every binary segmentation mask corresponding to a ROI in a fetal brain atlas is associated with an anatomical contract of trust.
Each of those binary segmentation masks represents the anatomy of a given tissue type, for a given gestational age and a given population of fetuses.
The anatomical contracts derived from atlas-based segmentation are therefore 
specific to a class, to a gestational age, and to the population of fetuses that was used to compute the atlas.
Since only neurotypical fetal brain atlases~\cite{gholipour2017normative,wu2021age} and a spina bifida fetal brain atlas~\cite{fidon2021atlas} are available in our work, our anatomical contract of trust will hold only for those two populations.

Due to the spatial smoothness imposed to the spatial transformation, atlas-based automatic segmentations will usually be correct up to a spatial margin.
Therefore, we propose to
to compute the BPAs of our anatomical contract of trust by
adding spatial margins to the atlas-based segmentation.
This approach is inspired by the safety margins used in radiotherapy to account for spatial registration errors~\cite{niyazi2016estro}.
%
%
Formally,
let $M^c$ a 3D (binary) mask from an atlas-based algorithm for class $c \in \mathbf{C}$.
We propose to define the BPA map $m^{(c)}=\left(m^{(c)}_{\textbf{x}}\right)_{\textbf{x}\in \Omega}$ associated with $M^c$ as
\begin{equation}
\label{eq:anatomical_bpa}
\forall \textbf{x},\quad
    \left\{
    \begin{aligned}
        m^{(c)}_{\textbf{x}}(\mathbf{C} \setminus \{c\}) &= 1 - \phi(d(\textbf{x}, M^c))\\
        m^{(c)}_{\textbf{x}}(\mathbf{C}) &=\phi(d(\textbf{x}, M^c))\\
    \end{aligned}
    \right.
\end{equation}
where $d(\textbf{x}, M^c)$ is the Euclidean distance from $\textbf{x}$ to $M^c$,
and $\phi: \mathbb{R}_{+} \xrightarrow{} [0,\, 1]$ with $\phi(0)=1$ and $\phi$ non-increasing.
\revnew{\revref{1}{3}We note that while $m^{(c)}$ is a BPA, it cannot be considered as a probability as it does not sum to one.}
In the following,
we use the function
\begin{equation}
    \forall d \geq 0,\quad \phi(d) = 
    \left\{
        \begin{array}{cc}
            1 & \texttt{if } d \leq \eta \\
            0 & \texttt{otherwise }
        \end{array}
    \right.
\end{equation}
where $\eta > 0$ is a hyper-parameter homogeneous to a distance and can be interpreted as a \textit{safety margin} for the anatomical prior.
We describe a method to tune the margins at training time for each class in appendix~\ref{sec:tuning_margins}.
The BPA for this function $\phi$ can be implemented efficiently without computing explicitly the distance between every voxel $\textbf{x}$ and the mask $M^c$.
%
With the definition of the BPA $m^{(c)}_{\textbf{x}}$ in \eqref{eq:anatomical_bpa}, we formalize the following belief: far enough from the mask $M^c$ we know for sure that the true class is not $c$, i.e. $m^{(c)}_{\textbf{x}}(\mathbf{C} \setminus \{c\}) = 1$, otherwise we do not know anything for sure regarding class $c$, i.e. $m^{(c)}_{\textbf{x}}(\mathbf{C}) > 0$.

The BPAs $m_c$ defined as in \eqref{eq:anatomical_bpa} are nowhere completely contradictory with each other. 
A proof can be found in Appendix \ref{appendix:proof-no-contradiction}.
Therefore, we can define the anatomical prior BPA
used in \eqref{eq:trustworhtyAI-fetal}
for image $I$ and voxel $\textbf{x}$ as
\begin{equation}
    \label{eq:anatomical_prior}
    m^{\anatomy}_{I, \textbf{x}} = \bigoplus_{c \in \mathbf{C}} m^{(c)}_{\textbf{x}}
\end{equation}
where $m_c$ is the BPA associated to the mask $M^c_I$ for class $c$ of the segmentation obtained using the multi-atlas fallback segmentation algorithm (see Appendix \ref{sec:fallback}).
We prove in Appendix \ref{appendix:proof-no-contradiction} that the proposed anatomical prior BPA is never completely contradictory with the fallback.

We prove that for all voxel $\textbf{x}$ and for all subset of classes $\mathbf{C}' \subset \mathbf{C}$,
the anatomical BPA mass that the true label of $\textbf{x}$ is not in $\mathbf{C}'$ is equal to
\begin{equation}
    \label{eq:anatomical_BPA}
    \begin{aligned}
         m^{\anatomy}_{I, \textbf{x}}(\mathbf{C}\setminus \mathbf{C}') =
    \prod_{c\in \mathbf{C}} &
    \left( 
        \delta_{c}(\mathbf{C}')m^{(c)}_{\textbf{x}}(\mathbf{C} \setminus \{c\}) \right.\\
      & \left. + (1 - \delta_{c}(\mathbf{C}'))m^{(c)}_{\textbf{x}}(\mathbf{C})
    \right)
    \end{aligned}
\end{equation}
where for all $c \in \mathbf{C}$, $\delta_c$ is the Dirac measure
defined as
\begin{equation}
    \forall \mathbf{C}' \subset \mathbf{C}, \quad \delta_c(\mathbf{C}') =
    \left\{
    \begin{array}{cc}
        1 & \textup{if}\,\, c \in \mathbf{C}'\\
        0 & \textup{if}\,\, c \not \in \mathbf{C}'
    \end{array}
    \right.
\end{equation}
The proof of \eqref{eq:anatomical_BPA} can be found in the Appendix.

In practice, we are particularly interested in summing the anatomical prior BPA with probabilities using the particular case of Dempster's rule in \eqref{eq:drc_proba}.
Let $\textbf{x}$ a voxel and $p_{I,\textbf{x}}$ a probability on $\mathbf{C}$ for voxel $\textbf{x}$ of image $I$ that is not completely contradictory with $m^{\anatomy}_{I,\textbf{x}}$.
For all $c \in \mathbf{C}$, we can show that
\begin{equation}
    \label{eq:DRC_proba_anatomical_BPA}
    \left(
    p_{I,\textbf{x}} \oplus m^{\anatomy}_{I,\textbf{x}}
    \right)
    \left(c\right) =
    \frac{p_{I,\textbf{x}}(c)m^{(c)}_{\textbf{x}}(\mathbf{C})}{\sum_{c'\in \mathbf{C}}p_{I,\textbf{x}}(c')m^{(c')}_{\textbf{x}}(\mathbf{C})}
\end{equation}
A proof of this equality can be found in the Appendix 
\revnew{and the pseudo-code can be found in Algorithm~\ref{alg:1}
}
.
It is worth noting that, due to the specific form of $m^{\anatomy}_{I,\textbf{x}}$ and because $p_{I,\textbf{x}}$ is a probability, the computational cost of $p_{I,\textbf{x}} \oplus m^{\anatomy}_{I,\textbf{x}}$ is $\mathcal{O}(\mathbf{C})$ even though there are $2^{|\mathbf{C}|}$ elements in $2^{\mathbf{C}}$.
Another important remark is that when $p_{I,\textbf{x}}$ is completely contradictory with $m^{\anatomy}_{I,\textbf{x}}$, we have
$\sum_{c'\in \mathbf{C}}p_{I,\textbf{x}}(c')m^{(c')}_{\textbf{x}}(\mathbf{C})=0$.

\begin{algorithm}[t!]\revnew{\revref{1}{3}}
\caption{
\revnew{
Dempster-Shafer anatomical contract of trust.
}
}
\label{alg:1}
\begin{algorithmic}[1]
\Require{$(p_{I, \textbf{x}})_{\textbf{x}\in \Omega}$: input class probability map.}
\Require{$(M^c)_{c \in \mathbf{C}}$: binary masks prior for all classes.}
\For{$c \in \mathbf{C}$}
    \State{$M^c \leftarrow Dilate_{\eta_{c}}(M^c)$}\Comment{Dilate mask by margin $\eta_{c}$}
    \State{$\forall \textbf{x} \in \Omega, \quad p_{I, \textbf{x}}(c) \leftarrow p_{I, \textbf{x}}(c) \times M^c_{\textbf{x}}$}\Comment{Mask}
\EndFor{}
\State{$\forall \textbf{x} \in \Omega, \forall c \in \mathbf{C} \quad 
p_{I, \textbf{x}} \leftarrow 
\frac{p_{I, \textbf{x}}(c)}{\sum_{c'}  p_{I, \textbf{x}}(c')}
$}\Comment{Normalize}
\State{\textbf{Output:} $(p_{I, \textbf{x}})_{\textbf{x}\in \Omega}$}
\end{algorithmic}
\end{algorithm}

\textbf{Tuning the margins:}
The margins $\eta$ were tuned for each class and each condition independently using the 3D MRIs of the fold 0 of the training dataset.
More details can be found in the appendix \ref{sec:tuning_margins}.



\subsubsection{Dempster-Shafer intensity-based contracts of trust}\label{sec:intensity_contract}
In this section, we describe our proposed intensity prior BPA $m^{\intensity}$ that is used in our trustworthy AI method for fetal brain 3D MRI segmentation \eqref{eq:trustworhtyAI-fetal}.

In T2-weighted MRI, hyper-intense voxels inside the brain are highly likely to be part of the cerebrospinal fluid (CSF).
Voxels outside the brain (\emph{background} class) can also be hyper-intense but not the non-CSF tissue types.
We therefore propose to model this intensity prior about high intensities as a contract of trust.
Regarding hypo-intense voxels, it is unclear how to derive similar prior because even the CSF classes contain hypo-intense voxels, such as the choroid plexus for the intra-axial CSF class and the vein of Galena and straight sinus for the extra-axial CSF class~\cite{payette2021automatic}.

Let $\mathbf{C}_{high}\subset \mathbf{C}$ be the subset of classes that contain all the classes that partition the entire CSF (intra-axial CSF and extra-axial CSF) and the background.
Let $I=\{I_{\textbf{x}}\}_{\textbf{x} \in \Omega}$ be the volume and $\Omega$ the volume domain of a fetal brain 3D MRI.
We propose to fit a Gaussian mixture model (GMM) with two components to the image intensity distribution of $I$.
The two components of parameters $(\mu_{high},\,\sigma_{high})$ and $(\mu_{low},\,\sigma_{low})$ are associated to high and low intensities.
We propose to define the intensity prior BPA for all voxels $\forall \textbf{x}$, up to a normalization factor, as
\begin{equation}
    \left\{
    \begin{aligned}
        m^{\intensity}_{I,\textbf{x}}(\mathbf{C}_{high}) &\propto \frac{1}{\sigma_{high}} \exp \left(
            \frac{1}{2} \left(\frac{I_{\textbf{x}} - \mu_{high}}{\sigma_{high}}\right)^2
        \right)\\
        m^{\intensity}_{I,\textbf{x}}(\mathbf{C}) &\propto \frac{1}{\sigma_{low}} \exp \left(
            \frac{1}{2} \left(\frac{I_{\textbf{x}} - \mu_{low}}{\sigma_{low}}\right)^2
        \right)\\
    \end{aligned}
    \right.
\end{equation}
It is worth noting that $m^{\intensity}_{I,\textbf{x}}(\mathbf{C}) > 0$.
Therefore, no probability will be set to $0$ using the Dempster's rule of combination with $m^{\intensity}$. In other words, $m^{\intensity}$ does not forbid any assignment.
This is in contrast with the anatomical BPAs defined in section~\ref{sec:anatomical_contract}.

Let $\textbf{x}$ a voxel and $p_{I,\textbf{x}}$ a probability on $\mathbf{C}$ for voxel $\textbf{x}$ of image $I$.
Since $m^{\intensity}_{I,\textbf{x}}(\mathbf{C}) >0$, $p_{I,\textbf{x}}$ is not completely contradictory with $m^{\intensity}_{I,\textbf{x}}$.
Using Dempster's rule, we have, for all class $c \in \mathbf{C}$
\begin{equation}
\begin{aligned}
    &\left(p_{I,\textbf{x}} \oplus m^{\intensity}_{I,\textbf{x}}\right)(c) \propto\\
    &\quad \left\{
    \begin{aligned}
        \left(
        1 + \frac{m^{\intensity}_{I,\textbf{x}}(\mathbf{C}_{high})}{m^{\intensity}_{I,\textbf{x}}(\mathbf{C})}
        \right) p_{I,\textbf{x}}(c) & \,\,\textup{if}\,\, c \in \mathbf{C}_{high}\\
        p_{I,\textbf{x}}(c) & \,\,\textup{otherwise}\\
    \end{aligned}
    \right.
\end{aligned}
\end{equation}
This can be interpreted as a soft-thresholding operation
Thus, only the probabilities for the background and CSF classes in $\mathbf{C}_{high}$ are increased in the case of a voxel $\textbf{x}$ with relatively high intensity.
In particular, the probabilities remain approximately unchanged  for a voxel $\textbf{x}$ with relatively low or medium intensity.
This reflects the fact that the background and CSF classes also contain hypo-intense voxels.
The hyper-intense voxels must be in $\mathbf{C}_{high}$ while we ca not say anything about hypo-intense voxels in general. There are hypo-intense voxels in every class.

\begin{table*}[t]
	\centering
	\caption{
    \revnew{
	    \textbf{Evaluation of our contracts of trust for different AI models.} Best values for each AI model are in \textbf{bold} and best values overall are \underline{underlined}.
	    IN: in-scanner distribution,
	    OUT: out-of-scanner distribution,
	    int.: intensity contract of trust,
	    anat.: anatomical contract of trust.
    }
	}
	\begin{tabularx}{\textwidth}{c c c *{12}{Y}}
		\toprule
        \multicolumn{1}{c}{Backbone AI} & \multicolumn{2}{c}{Contract}
        & \multicolumn{6}{c}{Mean-ROI Dice Score (in $\%$)} & \multicolumn{6}{c}{Mean-ROI HD95 (in mm)}\\
        %
        %
        \multicolumn{1}{c}{model} & \multicolumn{2}{c}{of trust}
        & \multicolumn{2}{c}{\bf Neurotypical} & \multicolumn{2}{c}{\bf Spina Bifida} & \multicolumn{2}{c}{\bf Other Path.}
		& \multicolumn{2}{c}{\bf Neurotypical} & \multicolumn{2}{c}{\bf Spina Bifida} & \multicolumn{2}{c}{\bf Other Path.}\\
        \cmidrule(lr){2-3} \cmidrule(lr){4-9} \cmidrule(lr){10-15}
		\multicolumn{1}{c}{}
		& \multicolumn{1}{c}{\bf int.} & \multicolumn{1}{c}{\bf anat.}
		& IN & OUT & IN & OUT & IN & OUT & IN & OUT & IN & OUT & IN & OUT \\
		\midrule
		Fallback
		& NA & NA
		& \mbox{85.7 (2.2)} & \mbox{84.1 (3.6)}
		& \mbox{78.8 (6.2)} & \mbox{76.2 (10)}
		& \mbox{78.6 (9.2)} & \mbox{82.5 (6.0)}
		& \mbox{1.5 (0.3)} & \mbox{1.5 (0.4)}
		& \mbox{\underline{2.3 (0.7)}} & \mbox{2.8 (1.6)}
		& \mbox{3.7 (2.6)} & \mbox{2.4 (1.6)} \\
	\cmidrule(lr){1-15}
	nnU-Net \cite{isensee2021nnu} 
		& \xmark & \xmark
		& \mbox{90.4 (1.8)} & \mbox{86.6 (3.8)}
		& \mbox{80.6 (6.9)} & \mbox{75.2 (14)}
		& \mbox{\bf 83.6 (8.7)} & \mbox{82.7 (5.6)}
		& \mbox{2.0 (0.4)} & \mbox{1.9 (0.4)}
		& \mbox{3.5 (1.7)} & \mbox{4.5 (2.9)}
		& \mbox{3.5 (2.3)} & \mbox{3.3 (2.5)} \\
        & \cmark & \xmark
		& \mbox{90.4 (1.8)} & \mbox{86.7 (3.9)}
		& \mbox{80.8 (6.9)} & \mbox{75.3 (14)}
		& \mbox{\bf 83.6 (8.6)} & \mbox{83.0 (5.0)}
		& \mbox{2.0 (0.4)} & \mbox{1.8 (0.4)}
		& \mbox{3.4 (1.7)} & \mbox{4.6 (2.9)}
		& \mbox{3.5 (2.3)} & \mbox{3.3 (2.4)} \\
        & \xmark & \cmark
		& \mbox{91.0 (1.7)} & \mbox{\bf 87.4 (3.7)}
		& \mbox{82.1 (5.8)} & \mbox{\bf 78.1 (10)}
		& \mbox{83.3 (9.7)} & \mbox{84.0 (5.7)}
		& \mbox{\bf 1.2 (0.2)} & \mbox{\underline{\bf 1.3 (0.3)}}
		& \mbox{\underline{\bf 2.3 (1.1)}} & \mbox{\bf 2.8 (1.6)}
		& \mbox{\underline{\bf 3.2 (2.6)}} & \mbox{2.1 (1.3)} \\
        & \cmark & \cmark
		& \mbox{\bf 91.1 (1.6)} & \mbox{\bf 87.4 (3.7)}
		& \mbox{\bf 82.2 (5.8)} & \mbox{78.0 (11)}
		& \mbox{83.3 (9.7)} & \mbox{\bf 84.2 (5.4)}
		& \mbox{\bf 1.2 (0.2)} & \mbox{\underline{\bf 1.3 (0.3)}}
		& \mbox{\underline{\bf 2.3 (1.1)}} & \mbox{\bf 2.8 (1.6)}
		& \mbox{\underline{\bf 3.2 (2.6)}} & \mbox{\underline{\bf 2.0 (1.3)}} \\
	\cmidrule(lr){1-15}
	SwinUNETR 
		& \xmark & \xmark
		& \mbox{84.3 (5.3)} & \mbox{77.9 (3.6)}
		& \mbox{74.2 (11)} & \mbox{65.2 (14)}
		& \mbox{79.9 (10)} & \mbox{79.2 (5.7)}
		& \mbox{2.9 (1.0)} & \mbox{3.1 (0.8)}
		& \mbox{5.3 (3.1)} & \mbox{6.6 (3.0)}
		& \mbox{4.9 (3.0)} & \mbox{6.1 (2.4)} \\
    with
        & \cmark & \xmark
		& \mbox{84.8 (5.1)} & \mbox{78.2 (3.6)}
		& \mbox{75.1 (10)} & \mbox{66.5 (14)}
		& \mbox{80.5 (10)} & \mbox{79.6 (5.6)}
		& \mbox{2.8 (1.0)} & \mbox{3.0 (0.8)}
		& \mbox{5.2 (3.1)} & \mbox{6.4 (3.1)}
		& \mbox{4.7 (2.8)} & \mbox{6.1 (2.5)} \\
	SSL
        & \xmark & \cmark
		& \mbox{86.5 (4.3)} & \mbox{80.7 (4.0)}
		& \mbox{78.4 (7.5)} & \mbox{70.9 (11)}
		& \mbox{81.1 (10)} & \mbox{81.1 (6.0)}
		& \mbox{1.4 (0.3)} & \mbox{1.7 (0.4)}
		& \mbox{\bf 2.4 (1.0)} & \mbox{3.1 (1.5)}
		& \mbox{\bf 3.3 (2.6)} & \mbox{\bf 2.2 (1.3)} \\
	pre-training~\cite{tang2022self}
        & \cmark & \cmark
		& \mbox{\bf 87.0 (3.9)} & \mbox{\bf 81.4 (3.9)}
		& \mbox{\bf 78.8 (7.3)} & \mbox{\bf 71.6 (11)}
		& \mbox{\bf 81.5 (10)} & \mbox{\bf 81.4 (6.0)}
		& \mbox{\bf 1.3 (0.2)} & \mbox{\bf 1.6 (0.3)}
		& \mbox{\bf 2.4 (1.0)} & \mbox{\bf 3.0 (1.5)}
		& \mbox{\bf 3.3 (2.6)} & \mbox{\bf 2.2 (1.2)} \\
	\cmidrule(lr){1-15}
	nnU-Net~\cite{isensee2021nnu} 
		& \xmark & \xmark
		& \mbox{90.5 (1.9)} & \mbox{86.6 (3.8)}
		& \mbox{80.4 (7.0)} & \mbox{76.8 (11)}
		& \mbox{83.4 (8.5)} & \mbox{82.5 (5.5)}
		& \mbox{1.9 (0.3)} & \mbox{1.8 (0.3)}
		& \mbox{3.5 (1.8)} & \mbox{3.9 (1.8)}
		& \mbox{3.5 (2.3)} & \mbox{2.7 (1.2)} \\
    with
        & \cmark & \xmark
		& \mbox{90.5 (1.9)} & \mbox{86.7 (3.8)}
		& \mbox{80.5 (7.0)} & \mbox{76.7 (12)}
		& \mbox{\bf 83.5 (8.4)} & \mbox{82.8 (4.9)}
		& \mbox{1.9 (0.3)} & \mbox{1.8 (0.3)}
		& \mbox{3.5 (1.7)} & \mbox{3.9 (1.8)}
		& \mbox{3.5 (2.3)} & \mbox{2.6 (1.1)} \\
	atlas features
        & \xmark & \cmark
		& \mbox{91.1 (1.7)} & \mbox{87.3 (3.7)}
		& \mbox{82.0 (5.9)} & \mbox{\bf 78.5 (10)}
		& \mbox{82.9 (9.8)} & \mbox{83.4 (6.2)}
		& \mbox{\bf 1.2 (0.2)} & \mbox{\underline{\bf 1.3 (0.3)}}
		& \mbox{\underline{\bf 2.3 (1.1)}} & \mbox{\underline{\bf 2.7 (1.5)}}
		& \mbox{\bf 3.3 (2.6)} & \mbox{\bf 2.1 (1.3)} \\
	fusion~\cite{liu2022single,kushibar2018automated}
        & \cmark & \cmark
		& \mbox{\underline{\bf 91.2 (1.6)}} & \mbox{\bf 87.4 (3.7)}
		& \mbox{\bf 82.1 (5.8)} & \mbox{78.4 (10)}
		& \mbox{82.9 (9.8)} & \mbox{\bf 83.6 (5.8)}
		& \mbox{\bf 1.2 (0.2)} & \mbox{\underline{\bf 1.3 (0.3)}}
		& \mbox{\underline{\bf 2.3 (1.1)}} & \mbox{\underline{\bf 2.7 (1.6)}}
		& \mbox{\bf 3.3 (2.6)} & \mbox{\bf 2.1 (1.2)} \\
	\cmidrule(lr){1-15}
	Ensemble
		& \xmark & \xmark
		& \mbox{90.9 (1.7)} & \mbox{87.7 (3.8)}
		& \mbox{81.6 (6.5)} & \mbox{77.9 (11)}
		& \mbox{83.5 (9.3)} & \mbox{84.6 (5.4)}
		& \mbox{1.6 (0.3)} & \mbox{1.5 (0.2)}
		& \mbox{3.1 (1.6)} & \mbox{3.4 (1.6)}
		& \mbox{3.4 (2.4)} & \mbox{2.5 (1.5)} \\
    nnU-Net~\cite{isensee2021nnu}
        & \cmark & \xmark
		& \mbox{91.0 (1.7)} & \mbox{87.7 (3.9)}
		& \mbox{81.7 (6.6)} & \mbox{77.9 (12)}
		& \mbox{\underline{\bf 83.9 (8.6)}} & \mbox{84.9 (4.8)}
		& \mbox{1.6 (0.2)} & \mbox{1.5 (0.2)}
		& \mbox{3.1 (1.6)} & \mbox{3.4 (1.6)}
		& \mbox{3.4 (2.5)} & \mbox{2.4 (1.5)} \\
	+
        & \xmark & \cmark
		& \mbox{\underline{\bf 91.2 (1.7)}} & \mbox{\underline{\bf 87.9 (3.8)}}
		& \mbox{\underline{\bf 82.5 (5.7)}} & \mbox{\underline{\bf 79.0 (10)}}
		& \mbox{83.2 (10)} & \mbox{84.8 (6.2)}
		& \mbox{\underline{\bf 1.1 (0.2)}} & \mbox{\underline{\bf 1.3 (0.3)}}
		& \mbox{\underline{\bf 2.3 (1.1)}} & \mbox{\underline{\bf 2.7 (1.6)}}
		& \mbox{\bf 3.3 (2.6)} & \mbox{\underline{\bf 2.0 (1.3)}} \\
	atlas
        & \cmark & \cmark
		& \mbox{\underline{\bf 91.2 (1.7)}} & \mbox{\underline{\bf 87.9 (3.8)}}
		& \mbox{\underline{\bf 82.5 (5.8)}} & \mbox{78.6 (11)}
		& \mbox{83.2 (9.9)} & \mbox{\underline{\bf 85.0 (5.7)}}
		& \mbox{\underline{\bf 1.1 (0.2)}} & \mbox{\underline{\bf 1.3 (0.3)}}
		& \mbox{\underline{\bf 2.3 (1.1)}} & \mbox{\underline{\bf 2.7 (1.6)}}
		& \mbox{\bf 3.3 (2.6)} & \mbox{\underline{\bf 2.0 (1.2)}} \\
	\bottomrule
	\end{tabularx}
	\label{tab:models_results}
\end{table*}

\section{Experiments}

\subsection{Evaluation on a large multi-center dataset.}
To effectively evaluate the performance of our trustworthy AI framework as a suitable method to improve the trustworthiness of a backbone AI model using a fallback model, we have selected the task of fetal brain segmentation in 3D MRI.
This task is clinically relevant and is characterized by large image protocol variations and large anatomical variations.

Deep learning-based AI methods for fetal brain MRI segmentation have recently defined state-of-the-art segmentation performance~\cite{fetit2020deep,fidon2021label,fidon2021partial,hong2020fetal,khalili2019automatic,li2021cas,payette2021automatic,zhao2022automated}, gradually replacing image registration-based segmentation methods~\cite{makropoulos2018review} in the literature.
Most previous work on deep learning for fetal brain MRI segmentation trained and evaluated their models using only MRIs of healthy fetuses or only MRIs acquired at one center.
However, 
the segmentation performance of deep learning methods typically degrades when images from a different center or a different scanner vendor as the one used for training are used or when evaluating the segmentation performance on abnormal anatomy~\cite{alis2021inter,kamraoui2022deeplesionbrain,maartensson2020reliability,oakden2020hidden,perone2019unsupervised,redko2019advances}.
One study has reported such issues for fetal brain MRI segmentation~\cite{fidon2021distributionally}.
Thus, 
\revmod{
we have evaluated the proposed trustworthy AI approach with four different backbone AI algorithms based on deep learning~\cite{isensee2021nnu,liu2022single,kushibar2018automated,tang2022self}
and a fallback algorithm consisting of a registration-based segmentation method~\cite{cardoso2015geodesic}.
Details of the backbone AI and fallback methods can be found in the appendix\ref{sec:nnunet}\ref{sec:other_backbone_AI}\ref{sec:fallback}.
}
%
We have used a large multi-centric fetal brain MRI dataset that consists of a total of $540$ 3D MRIs with neurotypical or abnormal brain development, with gestational ages ranging from $19$ weeks to $40$ weeks, and with MRIs acquired at $13$ hospitals across six countries.
%
%
The task consists of segmenting automatically a fetal brain 3D MRI into eight 
tissue types:
the corpus callosum, the white matter, the cortical gray matter, the deep gray matter, the cerebellum, the brainstem, the intra-axial cerebrospinal fluid (CSF), and the extra-axial CSF.

\subsection{Stratified evaluation across brain conditions and acquisition centers.}

\begin{figure}[t!]
    \centering
    \includegraphics[width=\linewidth,trim=1cm 0cm 0cm 0cm,clip]{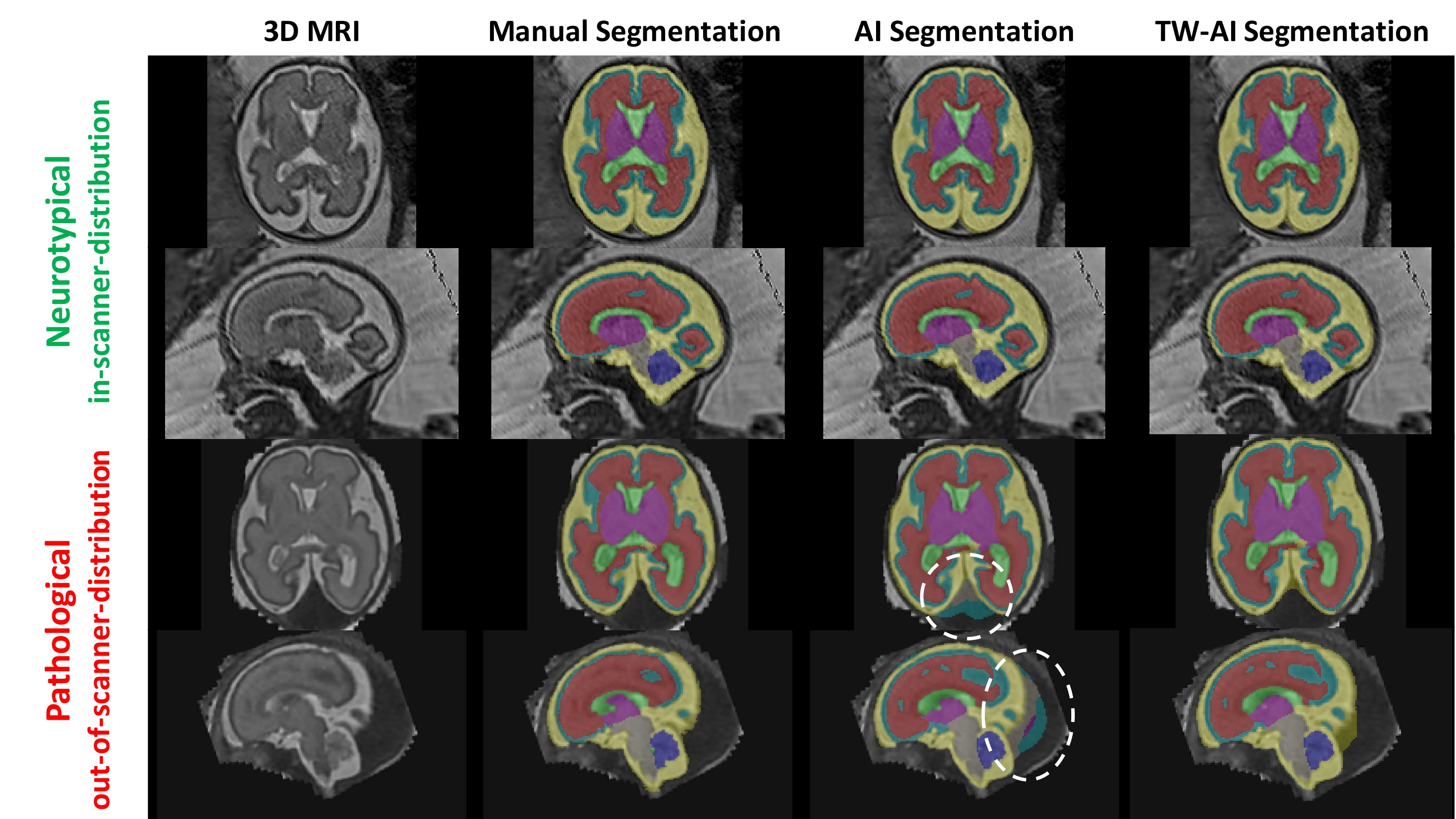}
    \caption{
        Illustration of the improved robustness of the proposed trustworthy AI method (TW-AI) as compared to \revmod{nnU-Net} state-of-the-art \revmod{backbone} AI method. (Top) 3D MRI of a neurotypical fetus at 28 weeks of gestation acquired at the same center as the training data for the AI. (Bottom) 3D MRI of a fetus with a high-flow dural sinus malformation at 28 weeks of gestation acquired at a different center as the training data for the AI. Severe violations of the anatomy by the \revmod{backbone} AI
        are highlighted.
        The TW-AI does not make those errors.
    }
    \label{fig:qualitative_results}
\end{figure}

The evaluation of AI-based segmentation algorithms has shown that the performance of deep learning models can vary widely across clinically relevant populations and across data acquisition protocols~\cite{fidon2021distributionally} (Fig.~\ref{fig:overview}b, Fig.~\ref{fig:qualitative_results}).

Therefore, we performed a stratified comparison of the backbone AI \revmod{algorithms}, the fallback algorithm, and the trustworthy AI \revmod{approach} across two groups of acquisition centers and three groups of brain conditions.
The composition of the dataset for each group is summarized in Fig.~\ref{fig:data} and detailed in section~\ref{sec:dataset}.
The acquisition centers were split into two groups, that we called \textit{in-scanner distribution} and \textit{out-of-scanner distribution}, depending if 3D MRIs acquired at a given center were present in the training dataset or not.
Four out of thirteen data sources were used 
\revmod{to train the backbone AI algorithms}.
In addition, the 3D MRIs were also separated based on the underlying brain condition of the fetus.
The first group, \textit{neurotypical}, contains the fetuses diagnosed by radiologists with a normal brain development using ultrasound and MRI.
The second group, \textit{spina bifida}, contains the fetuses with a condition called spina bifida aperta. We use the term \textit{spina bifida} for short in this work.
Cases of spina bifida aperta are typically accompanied by severe anatomical brain abnormalities~\cite{fidon2021atlas,pollenus2020impact}
with a type II Chiari malformation and an enlargement of the ventricles being most prevalent.
The Chiari malformation type II is characterized by a small posterior fossa and hindbrain herniation in which the medulla, cerebellum, and fourth ventricule are displaced caudally into the direction of the spinal canal~\cite{naidich1980computed}.
The third group, \textit{other pathologies}, contains fetuses with various pathologies other than spina bifida and causing an abnormal brain development, such as 
corpus callosum agenesis and dysgenesis,
intracranial hemorrhage and cyst,
aqueductal stenosis,
and Dandy-Walker malformation.
Those other pathologies were not present in the training dataset of the backbone AI \revmod{algorithms} and spatio-temporal atlases are not available for the fallback and the fail-safe algorithm.
Hence, testing 3D MRIs classified as other pathologies allow us to measure the segmentation performance of the trustworthy AI \revmod{approach} outside of the domain covered by the anatomical contracts of trust.
%
\revmod{Table~\ref{tab:models_results} shows}
the results of the overall stratified evaluation in terms of Dice score and Hausdorff distances at $95\%$ percentile 
\revnew{for four different backbone AI models.}
The detailed results per 
\revmod{ROI for nnU-Net as backbone AI}
can be found in the appendix (Fig.~\ref{fig:dice_roi},\ref{fig:hausdorff_roi}).
\revnew{Statistical differences were evaluated using a Wilcoxon singed-rank test 
using the threshold $0.05$ for the p-values.}

\revnew{
\subsection{Comparison with other fusion methods.}
In Table~\ref{tab:models_results}\revref{1}{4}, we compared our Dempster-Shafer trustworthy AI with nnU-Net as backbone AI, \emph{TW-nnU-Net} for short, to two other fusion methods based only on probabilities:
the ensemble average of the predicted probabilities of nnU-Net~\cite{isensee2021nnu} with the ones of the atlas-based fallback method~\cite{cardoso2015geodesic},
and \emph{nnU-Net with atlas features fusion}, in which the fusion operation between the atlas features and nnU-Net features is learned during training.
%
In this case, the atlas-based fallback probabilities are concatenated with the high-resolution deep features before the last level of the decoder~\cite{kushibar2018automated,liu2022single}.

We found that our TW-nnU-Net significantly outperformed nnU-Net with atlas features fusion for all groups and all metrics except in terms of Dice score for the in-scanner distribution other pathologies group for which they performed similarly.
Our TW-nnU-Net also significantly outperformed the ensemble average for all groups in terms of Hausdorff distance.
Our TW-nnU-Net and the ensemble average performed similarly in terms of Dice score.


}

\revnew{
\subsection{Ablation study of the proposed contracts of trust.}

The results in Table~\ref{tab:models_results}\revref{1}{5} show the benefits of both intensity and anatomical contracts of trust.
The anatomical contract of trust lead to similar or significantly better segmentation results for all backbone AI methods, all groups, and all metrics.\revref{2}{5}
We found that our intensity contract of trust applied to SwinUNETR~\cite{tang2022self} lead to significant improvement of the Dice score for the majority of the groups.
For the other backbone AI models, all based on nnU-Net~\cite{isensee2021nnu}, the segmentation metrics were similar for all groups with or without the use of the intensity contract of trusts.
We hypothesize that the various data augmentations used in nnU-Net, previously proposed in the domain generalization literature \cite{zhang2020generalizing}, allow those backbone AI models to learn robust intensity prior.
Similarly, we found that combining anatomical and intensity contracts of trust lead to significant improvement of the mean-ROI Dice score for SwinUNETR for the majority of the groups but to similar segmentation performance for the other backbone AI models.
Those results also show that our trustworthy AI approach can be successfully applied to various backbone AI models.

}

\subsection{Scoring of trustworthiness by radiologists.}
The Dice score and the Hausdorff distance are the two most standard metrics used for measuring the quality of automatic segmentations.
However, those two metrics do not directly measure the trustworthiness of segmentation algorithms~\cite{kofler2021we}.
Therefore, we have also conducted an evaluation of the trustworthiness of the automatic segmentations as perceived by radiologists.
We have asked a panel of \numraters{} experts from \numscoring{} different hospitals to score the trustworthiness of automatic segmentations from $0$ (totally unacceptable) to $5$ (perfect fit) for each region of interest and for the 
\revmod{
nnU-Net backbone AI, the fallback, and the corresponding trustworthy AI algorithms.
}
%
Independent scoring were performed by raters at different hospitals.
%
The scoring protocol and details about the panel of experts can be found in section~\ref{sec:scoring_protocol}.
The scoring was performed for the same 3D MRIs by all radiologists.
We have used a subset of $50$ 3D MRIs from the out-of-distribution group of the testing dataset consisting of $20$ neurotypical fetuses, $20$ spina bifida fetuses, and $10$ fetuses with other abnormalities.
Those cases were selected per condition at random among the 3D MRIs of the publicly available FeTA dataset~\cite{payette2021automatic}.
%
%
The \textit{out-of-scanner distribution} group is the most relevant group for the evaluation of trustworthiness because this corresponds to the situation in which AI algorithms generalization is the most challenging and clinically relevant.
%
The overall scoring results can be found in Fig.~\ref{fig:scores} and the detailed results per region of interest can be found in the appendix (Fig.~\ref{fig:scores_roi}).

\revnew{\revref{2}{4}
There were $7$ volumes out of $50$ for which the trustworthy AI approach achieved a lower average score than the backbone AI model.
However, with a maximum decrease of  $-0.32$ on our range of scores, we consider that there were no failure cases.
In contrast, there were $33$ volumes for which the trustworthy AI approach improved the backbone AI model average score by more than $0.32$, including $7$ volumes with an increase superior to $1$ and a maximum of $1.8$.
}

Expert raters 
noticed that the algorithms were dependent on the quality of the 3D MRIs they were based on for the \textit{spina bifida} group.
We found a positive correlation between the mean-class trustworthiness scores and the quality of the 3D MRI for the \textit{spina bifida} group (Pearson $r=0.43$).
There was no correlation between scores and 3D MRI quality for the \textit{neurotypical} group (Pearson $r=-0.1$) and the 3D MRIs of the \textit{other pathologies} were all of high quality.
%
%
In addition, the more structurally abnormal the brains were due to the pathologies, the more difficult it was to compare the algorithms. In the case of the Chiari malformations, this applies in particular to the cerebellum and brainstem.

\subsection{Stratified evaluation across gestational ages.}

The anatomy and the size of the fetal brain change significantly from $19$ weeks of gestation until term for both neurotypical fetuses~\cite{gholipour2017normative} and fetuses with spina bifida~\cite{fidon2021atlas}.
This age-related variability is a challenge for segmentation algorithms for fetal brain MRI~\cite{fidon2021distributionally}.
 
\begin{figure*}[t]
    \centering
    \subfloat[\textbf{Mean and $95\%$ CI of mean-ROI Dice Scores}]{\hspace{0.4cm}\includegraphics[width=0.4\textwidth,trim=0cm 0cm 0cm 3cm,clip]{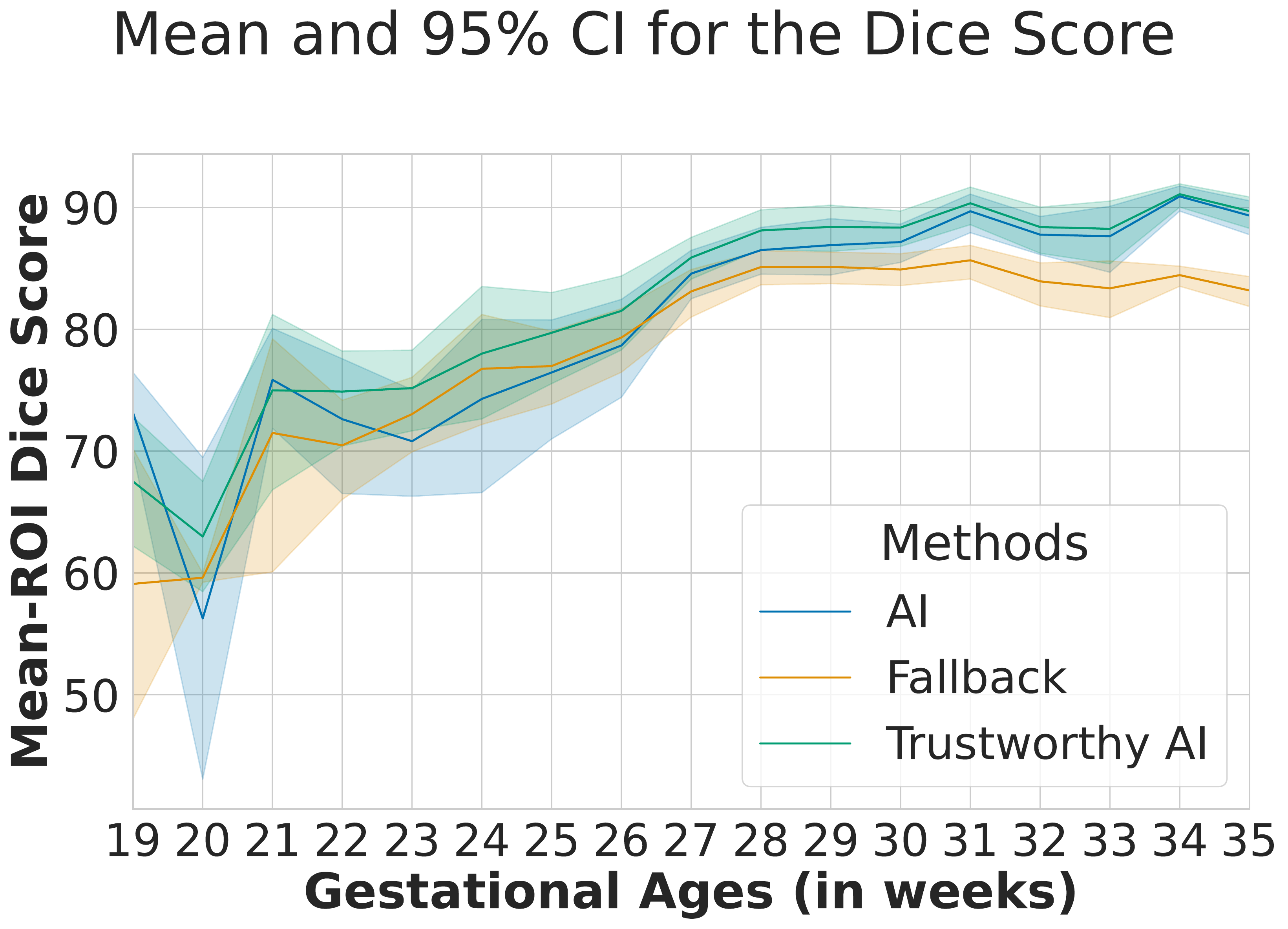}\label{fig:dice_GA}\hspace{0.4cm}}
    \subfloat[\textbf{Mean and $95\%$ CI of mean-ROI HD95}]{\hspace{0.4cm}\includegraphics[width=0.4\textwidth,trim=0cm 0cm 0cm 3cm,clip]{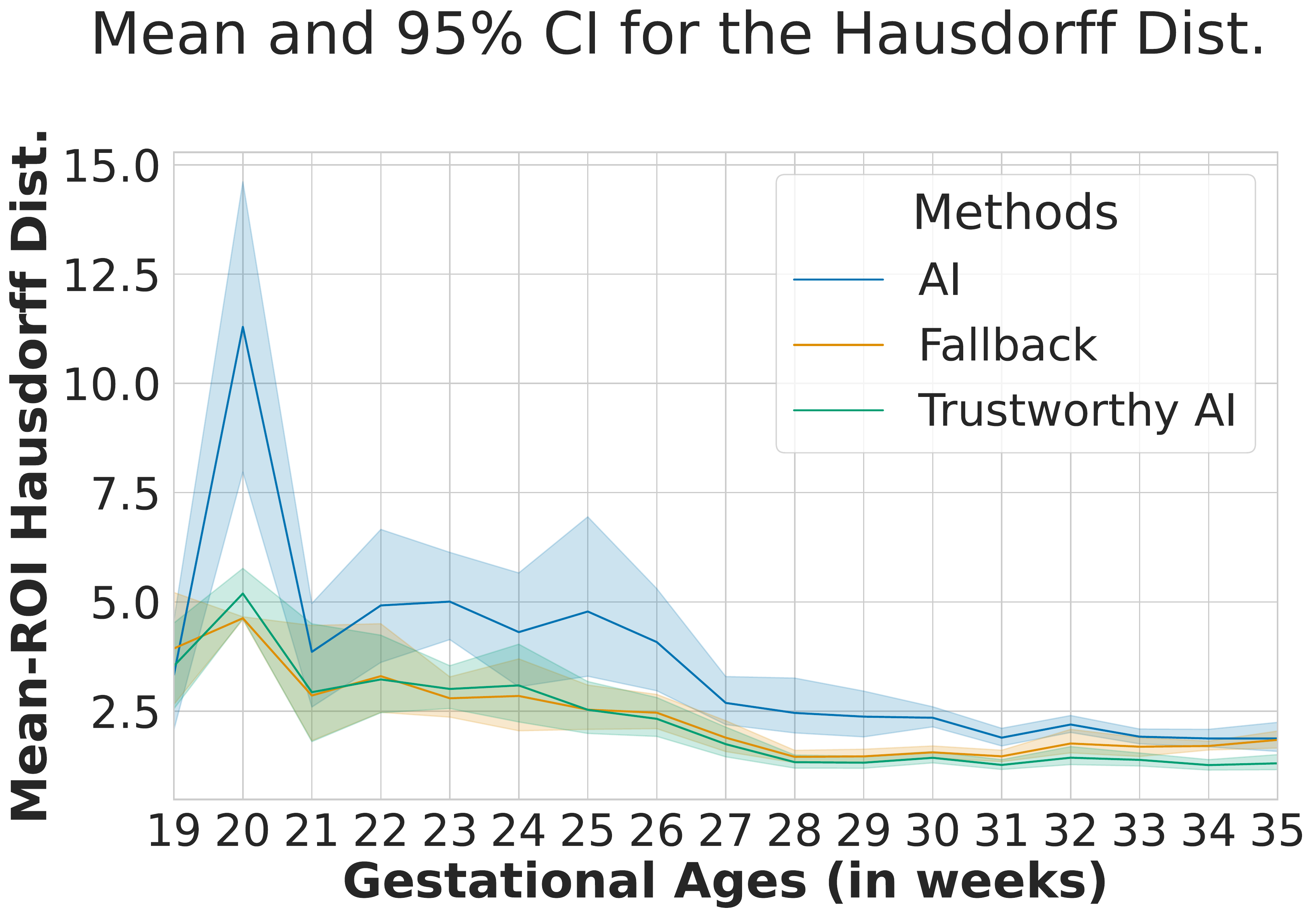}\label{fig:hausdorff_GA}\hspace{0.4cm}}\\
    \caption{
    \revmod{
    \textbf{
    Comparison of the backbone AI, fallback, and trustworthy AI segmentation algorithms across gestational ages, for neurotypical and spina bifida cases.
    }
    AI corresponds to nnU-Net~\cite{isensee2021nnu} here.
    Results per ROI can be found in the appendix (Fig.~\ref{fig:dice_GA_roi},~\ref{fig:hausdorff_GA_roi}).
    }
    }
    \label{fig:results}
\end{figure*}

We analysed the performance of the proposed trustworthy AI algorithm for fetal brain segmentation as a function of the gestational age and compared it to the 
\revmod{nnU-Net}\cite{isensee2021nnu}
backbone AI algorithm
and the fallback algorithm based on image registration~\cite{cardoso2015geodesic}.
We grouped the fetuses with neurotypical or spina bifida condition with the same gestational age rounded to the closest week.
The mean and the confidence intervals at $95\%$ for the overall performance in terms of Dice sore (resp. Hausdorff distance) across regions of interest can be found in Fig.~\ref{fig:dice_GA} (resp. Fig.~\ref{fig:hausdorff_GA}).
The detailed results per region of interest can be found in the appendix (Fig.~\ref{fig:dice_GA_roi},\ref{fig:hausdorff_GA_roi}).
%
Overall, the \revmod{nnU-Net} backbone AI algorithm achieves higher Dice scores than the fallback algorithm, while the fallback achieved lower Hausdorff distances than the backbone AI method (Fig.~\ref{fig:results}).
Our proposed trustworthy AI algorithm successfully combines backbone AI and fallback algorithms.
It achieves higher or similar segmentation performance than those two algorithms in terms of established segmentation quality metrics such as the Dice score and the Hausdorff distance across all gestational ages for \textit{neurotypical} and \textit{spina bifida}.
%



\begin{figure}[t!]
    \centering
    \includegraphics[width=0.98\linewidth,trim=0cm 0cm 0cm 3cm,clip]{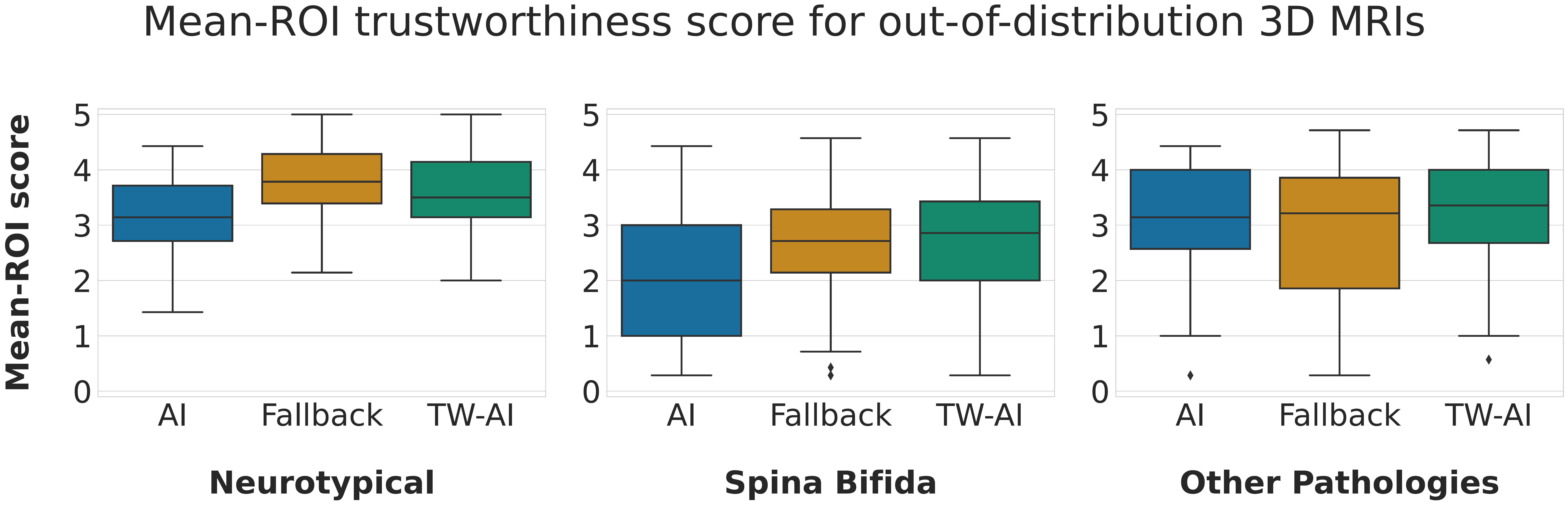}
    \caption{
    \textbf{Mean-ROI Trustworthiness Scores for out-of-scanner distribution 3D MRIs.}
    We report \numscoring{} scoring by a panel of \numraters{} experts of the trustworthiness of the automatic segmentations for a subset of the out-of-scanner distribution testing 3D MRIs ($n=50$).
    Each expert was asked to score from $0$ (totally unacceptable) to 5 (perfect fit) the trustworthiness of each ROI. The scores displayed here are averaged across ROIs.
    \revmod{AI corresponds to nnU-Net~\cite{isensee2021nnu} here.}
    %
    Results per ROI can be found in the appendix (Fig.~\ref{fig:scores_roi}).
    }
    \label{fig:scores}
\end{figure}

\section{Discussion and Conclusion}

\subsection{A principled and practical trustworthy AI method.}
We have mathematically formalized a method for trustworthy AI with a fallback based on Dempster-Shafer theory.
For application to fetal brain MRI segmentation, we have shown that our trustworthy AI method can be implemented using anatomy-based and intensity-based priors.
We have proposed to interpret those priors as contracts of trust in Human-AI trust theory.
Altogether, we showed that our principled trustworthy AI method improves the robustness and the trustworthiness of \revmod{four} state-of-the-art 
\revmod{
backbone AI algorithms}
\revnew{
and outperforms other fusion methods based on probabilities
} 
for fetal brain 3D MRI segmentation.

\subsection{Complementarity of AI and atlas-based algorithms.}
AI-based algorithms and registration-based algorithms have different error patterns.
In several situations we have found that the registration-based method tends to achieve better segmentation performance in terms of Hausdorff distance as compared to the AI-based method while the AI-based method achieved better segmentation performance in terms of Dice score.
We have found that the segmentation performance of the fallback algorithm decreases less than for the backbone AI \revmod{algorithms}, when comparing out-of-scanner distribution to in-scanner distribution for neurotypical and spina bifida fetal brain 3D MRIs.
In our scoring of trustworthiness on out-of-scanner distribution data, we have also found that
the fallback algorithm outperformed the \revmod{nnU-Net} backbone AI algorithm for neurotypical and spina bifida cases (Fig.~\ref{fig:scores}).
We think this is because the anatomical prior used by registration-based segmentation methods prevents mislabelling voxels far from the real anatomy.
In contrast, AI-based method are unconstrained and such errors can occur.
This is what we observe for the out-of-distribution cases displayed in Fig.~\ref{fig:overview}c,~\ref{fig:qualitative_results}.
Our proposed fail-safe method uses the registration-based segmentation with added margins with the aim to automatically detect and discard such errors that were found to occur more often for AI-based approach than for registration-based approach.

\subsection{The contracts of trust hold for sub-populations covered by brain atlases.}
Our implementation of trustworthy AI for fetal brain segmentation depends on the availability of spatio-temporal segmentation atlases of the fetal brain in 3D MRI.
While such atlases currently exist for neurotypical fetal brain~\cite{gholipour2017normative,wu2021age} and fetuses with spina bifida~\cite{fidon2021atlas}, it is not the case for other fetal brain pathologies.
Therefore, our contracts of trust are not expected to hold for the group \textit{other pathologies}.
This illustrates how AI trustworthiness is context-dependent.
We found that the \textit{other pathologies} group is the only one for which radiologists associated the fallback method, based solely on the atlases, with a lower trustworthy scores than the backbone AI algorithm (Fig.~\ref{fig:scores}).
Surprisingly, we found that the trustworthy AI algorithm still performs better or on a par with the backbone AI algorithm for the \textit{other pathologies} group.
We associate this with the use of our margins and to the proposed voxel intensity prior for the cerebrospinal fluid that are specific to the trustworthy AI algorithm.
For the \textit{other pathologies} group, we used the margin values estimated for spina bifida. 
Our group \textit{other pathologies} gathers diverse rare developmental diseases associated with different variations of the fetal brain anatomy.
However, due to the low number of examinations available per pathology, grouping them was necessary for evaluation purposes.
This introduces biases when comparing the segmentation performance for the \textit{other pathologies} groups associated with in-scanner and out-of-scanner distribution.
In particular, some of the fetuses with \textit{other pathologies} in the in-scanner distribution had very severe brain anatomical abnormalities, such as acqueductal stenosis with large supratentorial ventricles and caudal discplacement of the cerebellum or intracranial hemorrhage with parenchymal destruction and ventriculomegaly.
In contrast, the one in the out-of-scanner distribution have milder brain abnormalities, such as moderate ventriculomegaly, and there were no cases with parenchymal destruction.
This explains why, for this condition, we observe more outliers with low Dice scores and high Hausdorff distances for the backbone AI \revmod{algorithms} for in-scanner-distribution as compared to out-of-distribution 3D MRIs 
\revmod{
(Table~\ref{tab:models_results}).
}
This is also the only group for which the trustworthy AI method does not significantly outperform the
backbone AI methods
in terms of Dice score.

The two histograms of gestational ages for the training spina bifida 3D MRIs and the in-scanner-distribution testing spina bifida 3D MRIs are not uniform and have the same shape (see Fig.~\ref{fig:data}).
In contrast, the histogram of gestational ages for the out-of-scanner-distribution testing spina bifida is more uniform.
This might partly explain the degradation of Dice scores and Hausdorff distances
between in-scanner-distribution and out-of-scanner-distribution for the backbone AI algorithms 
\revmod{
(Table~\ref{tab:models_results}).
}
Training and in-scanner-distribution testing spina bifida MRIs were mostly clinical data acquired at \uzlshort{}.
In this center, MRI of spina bifida are typically performed a few days before and after the surgery that is performed prior to $26$ weeks of gestation.
In addition, a follow-up MRI is sometimes performed one month after the surgery.
This explains the two modes observed in the histograms for those two groups.
In the training data, the use of the spina bifida atlas~\cite{fidon2021atlas}, that has a uniform gestational age distribution, makes the second mode less visible.
Our results suggest the trustworthy AI algorithm is more robust than the AI algorithm to the gestational ages distributional shift between training and testing.

For gestational ages lower than $27$ weeks, the Dice scores and Hausdorff distances degrade for all the algorithms (Fig.~\ref{fig:dice_GA},\ref{fig:hausdorff_GA}).
For the \revmod{nnU-Net} backbone AI this is surprising given that more MRIs acquired at gestational ages lower than $27$ weeks than higher were present in the training dataset (Fig.~\ref{fig:data}).
Poorer MRI quality, which is typical for younger fetuses, might explain this degradation.
In addition, the ratio of spina bifida over neurotypical examinations is higher for gestational ages lower than $27$ weeks in our dataset.
The abnormal brain anatomy of spina bifida cases leads to more difficult segmentation compared to neurotypical cases.
This is particularly the case for several classes: the cerebellum, the extra-axial cerebrospinal fluid (CSF), the cortical gray matter, and the brainstem (Fig.~\ref{fig:dice_roi},\ref{fig:hausdorff_roi},\ref{fig:dice_GA_roi},\ref{fig:hausdorff_GA_roi},\ref{fig:scores_roi}).
The cerebellum is more difficult to detect using MRI before surgery as compared to early or late after surgery~\cite{aertsen2019reliability,danzer2007fetal}.
This has already been found to affect the segmentation performance of AI-based algorithms~\cite{fidon2021distributionally}.
For neurotypical fetuses, the extra-axial CSF is present all around the cortex.
However, for fetal brain MRI of spina bifida fetuses with gestational ages of $27$ weeks or less this is often not the case and the extra-axial CSF might be reduced to several small connected components that do not embrace the entire cortex anymore.
The spina bifida atlas does not cover well this variability of the extra-axial CSF~\cite{fidon2021atlas}.
Due to the explicit spatial regularization, medical image registration cannot tackle such differences of topology.
Therefore, using the atlas currently available, the contract of trust for extra-axial CSF does not apply for this group of spina bifida cases.
It can also influence nearby regions, such as the cortical gray matter in this case.
For the fallback algorithm and the trustworthy AI algorithm, a further degradation of the segmentation performance for gestational ages lower than $21$ weeks was expected because the fetal brain atlases used start at $21$ weeks.
For gestational ages of $21$ weeks or higher, the trustworthy AI outperforms either the backbone AI-algorithm or the fallback algorithm and performs better or on a par with the best other algorithms for all regions of interest in terms of Dice score and Hausdorff distance (Fig.~\ref{fig:dice_GA_roi},\ref{fig:hausdorff_GA_roi}).
The confidence intervals are also similar or narrower for the trustworthy AI algorithm than for the other algorithms for gestational ages higher or equal to $21$ weeks.
This illustrates that our contracts of trust improve the robustness of the proposed trustworthy AI algorithm for spina bifida for the range of gestational ages covered by the atlas used~\cite{fidon2021atlas}.

\subsection{Future work.}
For this work we have created the largest manually segmented fetal brain MRI dataset to date that consists of $540$ fetal brain 3D MRIs from $13$ acquisition centers.
A recent trend in medical image processing using AI is to gather even larger multi-institutional datasets using methods such as federated learning~\cite{rieke2020future}.
One can hypothesize that, with enough data, the AI algorithm would get more accurate even in the worst case until eventually reaching the same accuracy as the trustworthy AI algorithm in all cases.
However, results of our stratified evaluation suggest that this will require manually annotated 3D MRIs for every scanner acquisition protocol, for every condition, and for every gestational age.
To give an order of magnitude of the required dataset size, if we consider that $10$ 3D MRIs are required for each gestational age from $19$ weeks to $38$ weeks, for each of $10$ conditions and each of $5$ hospitals, we would already need $10,000$ 3D MRIs for both training and testing.
Given the low prevalence of some conditions~\cite{fidon2021distributionally} and the cost of obtaining fully-segmented data, classical supervised learning approaches might not be sufficient.
This rough estimation does not even include important confounding factors such as ethnicity and gender.
Altogether, this suggests that gathering more training data to improve the AI algorithm prior to deployment might not be sufficient to make the AI algorithm alone trustworthy.

The proposed trustworthy AI approach is not limited to fetal brain MRI and we expect it to be applicable to many medical image segmentation problems.
The proposed fail-safe mechanism, that is part of our trustworthy AI method, could be used to help improving the backbone AI continuously after its deployment.
An AI incident could be declared when a large part of the AI algorithm prediction was discarded by the \emph{fail-safe mechanism}.
This would allow automatic detection of images to correct and include in priority in the training set to update the backbone AI algorithm.
In addition, reporting such incidents could help to further improve the trust of the user.
In the context of trust, it is important to report such issues even when the incidents were handled correctly using the fallback segmentation algorithm.
In addition, as part of the  European Union Medical Device Regulations (EU MDR) Article 87~\cite{mdr},
it is a requirement for medical device manufacturers to report device-related incidents.
Previous methods for global segmentation failures detection, i.e. at the image-level, were proposed~\cite{kofler2021robust,robinson2019automated}.
In contrast, our fail-safe mechanism approaches the problem locally, i.e. at the voxel-level, by using atlas-based and intensity-based priors.

The margins used in our trustworthy AI segmentation algorithm 
could also support interactive segmentation.
Instead of providing voxel-level corrections or scribbles, the annotator could interact with the automatic segmentation by manually adapting the margins for its annotation.
After manual adjustment, the voxels outside the margins are automatically marked as correctly labelled while for the voxel inside the margins will be assigned a set of possible labels.
This yields partial annotations that can be exploited to improve the backbone AI method using partially-supervised learning methods~\cite{fidon2021label}.
This use of margins is similar, in terms of user interaction, to the safety margins that are used in clinics for radiation therapy planning~\cite{niyazi2016estro}.

The expert raters also emphasized that some frequent major violations in the cortex layer could be quickly removed manually and that they would have given higher scores to the segmentations if they could interact with them.
This echoes previous work on computational-aided decision making that found that users are more satisfied with imperfect algorithms if they can interact with them~\cite{dietvorst2018overcoming}.
Our findings suggest that allowing interactions would also increase the trust in AI algorithms for medical image segmentation.

%



\ifCLASSOPTIONcompsoc
  \section*{Acknowledgments}
\else
  \section*{Acknowledgment}
\fi
This project has received funding from the European Union's Horizon 2020 research and innovation program under the Marie Sk{\l}odowska-Curie grant agreement TRABIT No 765148.
%
%
Tom Vercauteren is supported by a Medtronic / RAEng Research Chair [RCSRF1819\textbackslash7\textbackslash34].
%

\ifCLASSOPTIONcaptionsoff
  \newpage
\fi

\bibliographystyle{IEEEtran}
\bibliography{IEEEabrv,main}

\FloatBarrier
\cleardoublepage
\appendices

\renewcommand\thefigure{\thesection.\arabic{figure}}  
\renewcommand\thetable{\thesection.\arabic{table}}    
\renewcommand\theequation{\thesection.\arabic{equation}}    
\setcounter{figure}{0}
\setcounter{table}{0}
\setcounter{equation}{0}

\section{}

\subsection{Fetal brain MRI dataset}\label{sec:dataset}

\begin{figure}[!htb]
    \centering
    \includegraphics[width=\linewidth,trim=0cm 0cm 0cm 4cm,clip]{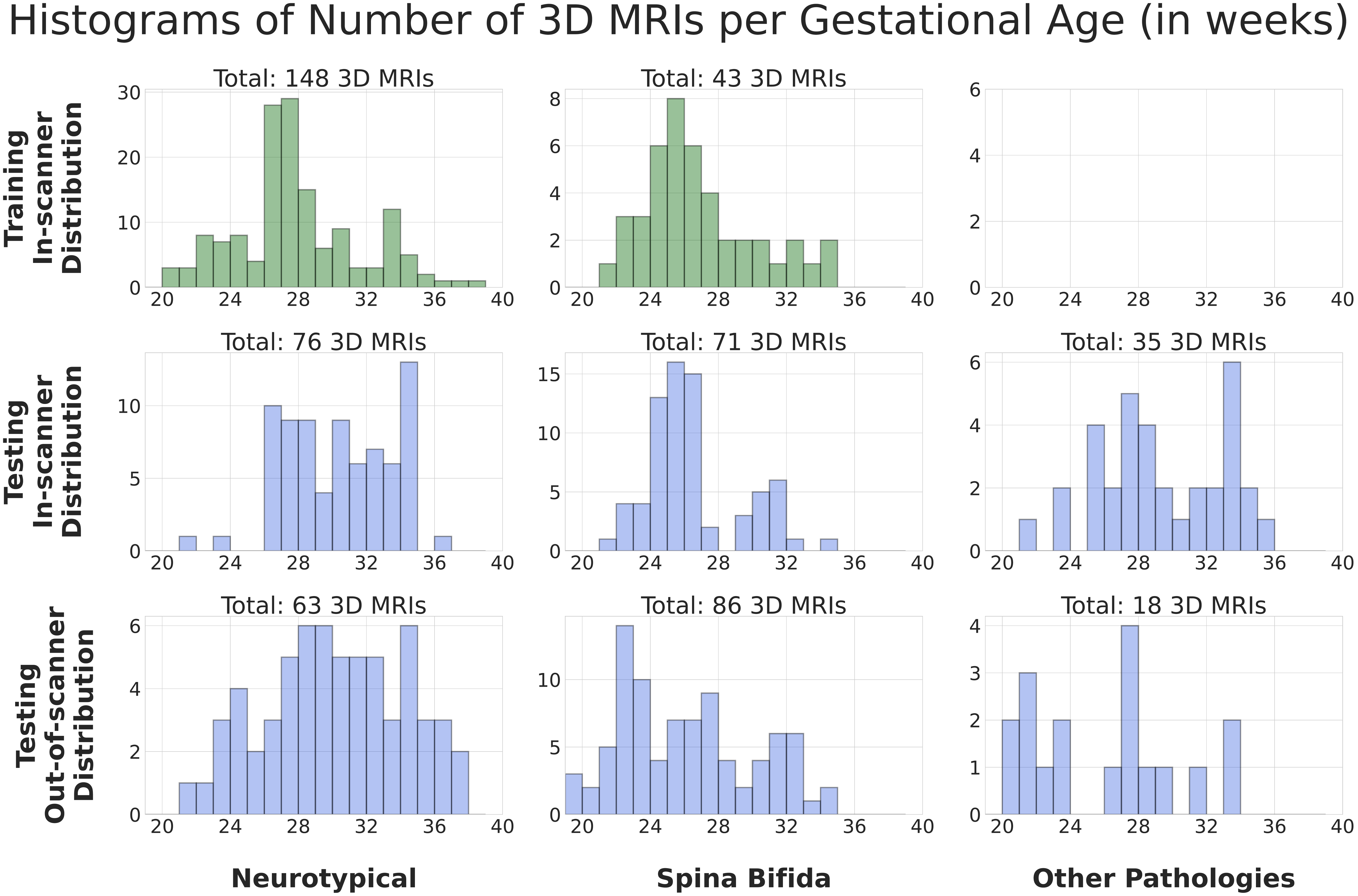}
    \caption{
    \textbf{Composition of the training and testing datasets (total: 540 3D MRIs)}
    \textit{In-scanner distribution} designates the 3D MRIs acquired at the same center as the training data.
    \textit{Out-of-scanner distribution} designates the 3D MRIs acquired at different centers than the training data.
    This is the largest fetal brain MRI dataset reported to date.
    }
    \label{fig:data}
\end{figure}

We have collected a dataset with a total of $540$ fetal brain 3D MRIs with neurotypical or abnormal brain development and from $13$ sources of data across $6$ countries.

The dataset consists of $326$ 3D MRIs acquired at \uzl{} (\uzlshort{}),
$88$ 3D MRIs from the FeTA dataset~\citesupp{payette2021automatic} (data release 1 and 2),
$11$ 3D MRIs acquired at \vienna{} (\viennashort{}),
$29$ 3D MRIs acquired at \kcl{} (\kclshort{}),
$27$ 3D MRIs acquired at \uclh{} (\uclhshort{}),
$4$ 3D MRIs acquired at \manchester{} (\manchestershort{}),
$4$ 3D MRIs acquired at \belfast{} (\belfastshort{}),
$2$ 3D MRIs acquired at \cork{} (\corkshort{}),
$1$ 3D MRIs acquired at \newcastle{} (\newcastleshort{}),
$1$ 3D MRIs acquired at \liverpool{} (\liverpoolshort{}),
and $47$ 3D MRIs from three fetal brain brain atlases.
The three open-access fetal brain spatio-temporal atlases consist of
$18$ population-averaged 3D MRIs computed from fetal neurotypical brain MRIs acquired at Boston Children's Hospital, USA~\citesupp{gholipour2017normative},
$14$ population-averaged 3D MRIs computed from fetal neurotypical brain MRIs acquired in China~\citesupp{wu2021age},
and $15$ population-averaged 3D MRIs computed from fetal spina bifida brain MRIs acquired at \uzlshort{} and \uclhshort{}~\citesupp{fidon2021atlas}.

Data from \uzlshort{} includes $192$ 3D MRIs of neurotypical fetuses, $99$ 3D MRIs of fetuses with spina bifida aperta, and $35$ 3D MRIs of fetuses with an abnormal brain anatomy due to a condition other than spina bifida.
The majority of the neurotypical fetuses was scanned for a suspected abnormality somewhere else than in the brain, while a minority was scanned for screening of brain abnormality but was proven neurotypical after MRI.
The $35$ 3D MRIs of fetuses with other abnormalities consisted of:
$3$ examinations of a case with an enlarged subarachnoid space,
$3$ cases of intraventricular hemorrhage,
$1$ cases of intracranial hemorrhage,
$1$ case with a partial rombencephalosynapsis,
$1$ case with a closed lip Schizencephaly,
$4$ cases with Dandy-Walker malformation,
$1$ case with an unilateral ventriculomegaly due to a hemorrhage,
$1$ case with choroid plexus papilloma,
$1$ case with high-flow dural sinus malformation,
$7$ cases with corpus callosum agnesis,
$1$ case with corpus callosum agenesis with interhemispheric cyst, temporal cysts and delayed gyration,
$2$ cases with tuberous sclerosis,
$1$ case with a Blake's pouch cyst,
$2$ cases with aqueductal stenosis,
$1$ case with an idiopathic dilatation of the lateral ventricles,
$2$ cases with cytomegalovirus encephalitis,
and $1$ case with parenchyma loss due to an ischemic insult.

Data from the publicly available FeTA dataset~\citesupp{payette2021automatic} includes $34$ 3D MRIs of fetuses with a normal brain development, $36$ 3D MRIs of fetuses with spina bifida aperta, and $18$ 3D MRIs of fetuses with an abnormal brain anatomy due to conditions other than spina bifida.
Those $18$ 3D MRIs of fetuses with other abnormalities consisted of: 
$3$ cases with heterotopia, 
$8$ cases with ventriculomegaly without spina bifida, 
$2$ cases with aqueductal stenosis, 
$2$ cases with interhemispheric cyst, 
$1$ case with cerebellar hemorrhage, 
$1$ case with a high-flow dural sinus malformation, and 
$1$ case with bilateral subependymal cysts and temporal cysts.

Data from \kclshort{} consists exclusively of brain 3D MRIs of fetuses with a normal brain development.
Data from \viennashort{}, \uclhshort{}, \manchestershort{}, \belfastshort{}, \corkshort{}, \newcastleshort{}, and \liverpoolshort{}
consist only of 3D MRIs of fetuses diagnosed with spina bifida aperta.

The composition of the training and testing datasets is summarized in Fig.\ref{fig:data}.
The training dataset consists of the $47$ volumes from the three fetal brain atlases, $144$ neurotypical cases from \uzlshort{} and $28$ spina bifida cases from \uzlshort{}.
The rest of the data is used for testing ($n=349$).
In the testing dataset, 3D MRIs that were acquired at \uzlshort{} are designated as \textit{in-scanner-distribution} while the data from other acquisition centers are designated as \textit{out-of-scanner-distribution} data.

The 3D MRIs with an isotropic image resolution of $0.8$mm have been reconstructed from the stacks of 2D MRI slices acquired at \uzlshort{}, \viennashort{}, \uclhshort{}, \manchestershort{}, \belfastshort{}, \corkshort{}, \newcastleshort{}, and \liverpoolshort{} using the state-of-the-art and publicly available software \href{https://github.com/gift-surg/NiftyMIC/tree/master/data/templates}{\texttt{NiftyMIC}}~\citesupp{ebner2020automated}.
The original 2D MRI slices were also corrected for bias field in the \href{https://github.com/gift-surg/NiftyMIC}{\texttt{NiftyMIC}} pipeline version $0.8$ using a N4 bias field correction step as implemented in \href{https://simpleitk.org}{SimpleITK} version $1.2.4$.
Brain masks for those MRIs were computed using \texttt{\href{https://github.com/gift-surg/MONAIfbs}{MONAIfbs}}~\citesupp{ranzini2021monaifbs}, an automatic method for fetal brain extraction in 2D fetal MRIs.
The 3D brain masks are reconstructed using the automatic 2D brain masks along with the 3D MRIs in \texttt{NiftyMIC}.

The 3D MRI reconstructions for the FeTA dataset is described in~\citesupp{payette2021automatic}.
Two volumes of spina bifida cases were excluded from the total FeTA dataset because the poor quality of the 3D MRI reconstruction (\texttt{sub-feta007} and \texttt{sub-feta009}) did not allow to manually segment them reliably for the seven tissue types.
The brain masks for those 3D MRIs were computed directly using the 3D MRIs and an atlas-based method as described in our previous work~\citesupp{fidon2021partial}.

\subsection{Human expert scoring method for evaluating the trustworthiness of fetal brain 3D MRI segmentation}\label{sec:scoring_protocol}
The trustworthiness scoring is done for each of the tissue types: white matter, intra-axial CSF, cerebellum, extra-axial CSF, cortical gray matter, deep gray matter, and brainstem.

The evaluation is performed using a Likert scale ranging from $0$ star to $5$ stars to answer the question \textit{"Is the automatic segmentation of the tissue type X trustworthy?"}:
\begin{description}
    \item[\faStarO\faStarO\faStarO\faStarO\faStarO] Strongly disagree / there are several severe violations of the anatomy that are totally unacceptable
    \item[\faStar\faStarO\faStarO\faStarO\faStarO] Disagree / there is one severe violation of the anatomy that is totally unacceptable
    \item[\faStar\faStar\faStarO\faStarO\faStarO] Moderately disagree / there are violations of the anatomy that make the acceptability of the segmentation questionable
    \item[\faStar\faStar\faStar\faStarO\faStarO] Moderately agree / there are many minor violations of the anatomy that are acceptable
    \item[\faStar\faStar\faStar\faStar\faStarO] Agree / there are a few minor violations of the anatomy that are acceptable
    \item[\faStar\faStar\faStar\faStar\faStar] Strongly agree / perfect fit of the anatomy
\end{description}

This evaluation is performed on $50$ 3D MRIs from the FeTA dataset. We selected at random $20$ neurotypical cases, $20$ spina bifida cases, and $10$ cases with other brain pathologies.

The scoring was performed independently by \numscoring{} individuals or groups of expert raters: 
MA, paediatric radiologist at \uzl{} with several years of experience in manual segmentation of fetal brain MRI;
AJ, MD and group leader at University Children’s Hospital Zurich;
AB, professor of neuroradiology at University Hospital Zurich;
and by MS and PK jointly, two MDs at \vienna{} (\viennashort{}) with more than $400$ hours of experience in manual segmentation of fetal brain MRI,
under the supervision of 3 experts:
DP, professor of radiology at \viennashort{},
GK, professor of paediatric radiology at \viennashort{},
and IP, neuroradiologist at \viennashort{}.

The human expert raters were given access to the 3D MRIs, the backbone AI segmentation, the fallback segmentation, the trustworthy AI segmentation, and the ground-truth manual segmentation for each case.
The segmentation algorithms are anonymized for the raters by assigning to each of them a number from 1 to 3.
The assignment of numbers to the segmentation methods was performed randomly for each case independently, i.e. we used a different random assignment for each case.

The ground-truth manual segmentations were not scored as border artefacts specific to manual segmentation made them impossible to anonymize them from the automatic segmentations.
However, we note that the trustworthiness of manual segmentation is likely to be impacted by the suboptimal image quality and abnornal brain anatomy.

%
\begin{figure}[!htb]
    \centering
    \includegraphics[width=\linewidth,trim=0cm 0cm 0cm 3.8cm,clip]{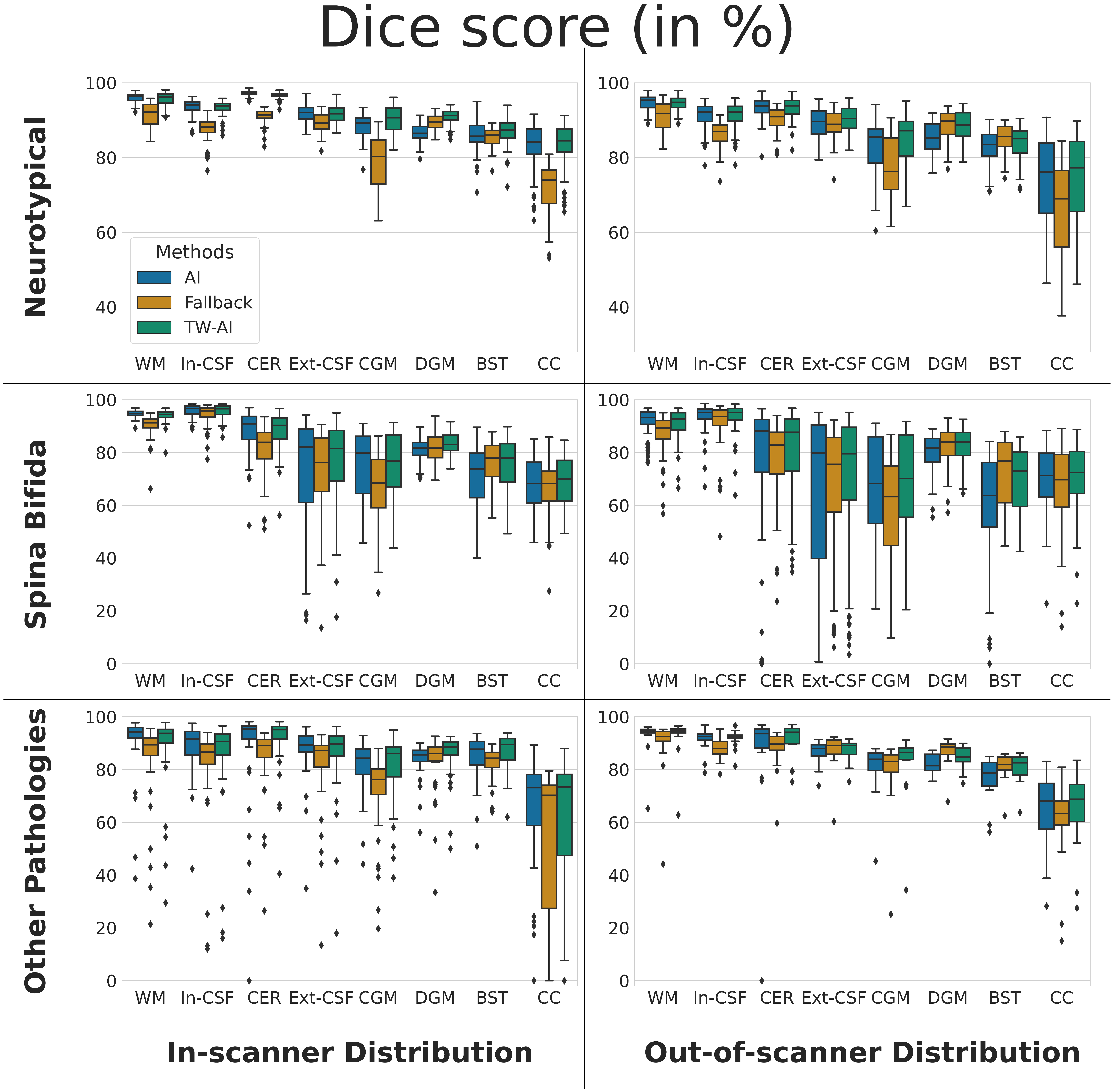}
    \caption{
    \textbf{Dice score (in \%) comparison of our AI, fallback, and trustworthy AI segmentation algorithms for fetal brain 3D MRI segmentation.}
    Dice scores are reported for all 3D MRI for $7$ tissue types:
    white matter (\textbf{WM}),
    intra-axial cerebrospinal fluid (\textbf{in-CSF}),
    cerebellum (\textbf{Cer}),
    extra-axial cerebrospinal fluid (\textbf{Ext-CSF}),
    cortical gray matter (\textbf{CGM}),
    deep gray matter (\textbf{DGM}),
    brainstem (\textbf{BST}),
    and corpus callosum (\textbf{CC}).
    Box limits are the first quartiles and third quartiles. The central ticks are the median values. The whiskers extend the boxes to show the rest of the distribution, except for points that are determined to be outliers.
    Outliers are data points outside the range median $\pm 1.5\times$ interquartile range.
    \revmod{AI corresponds to nnU-Net~\cite{isensee2021nnu} here.}
    }
    \label{fig:dice_roi}
\end{figure}

\begin{figure}[!htb]
    \centering
    \includegraphics[width=\linewidth,trim=0cm 0cm 0cm 3.8cm,clip]{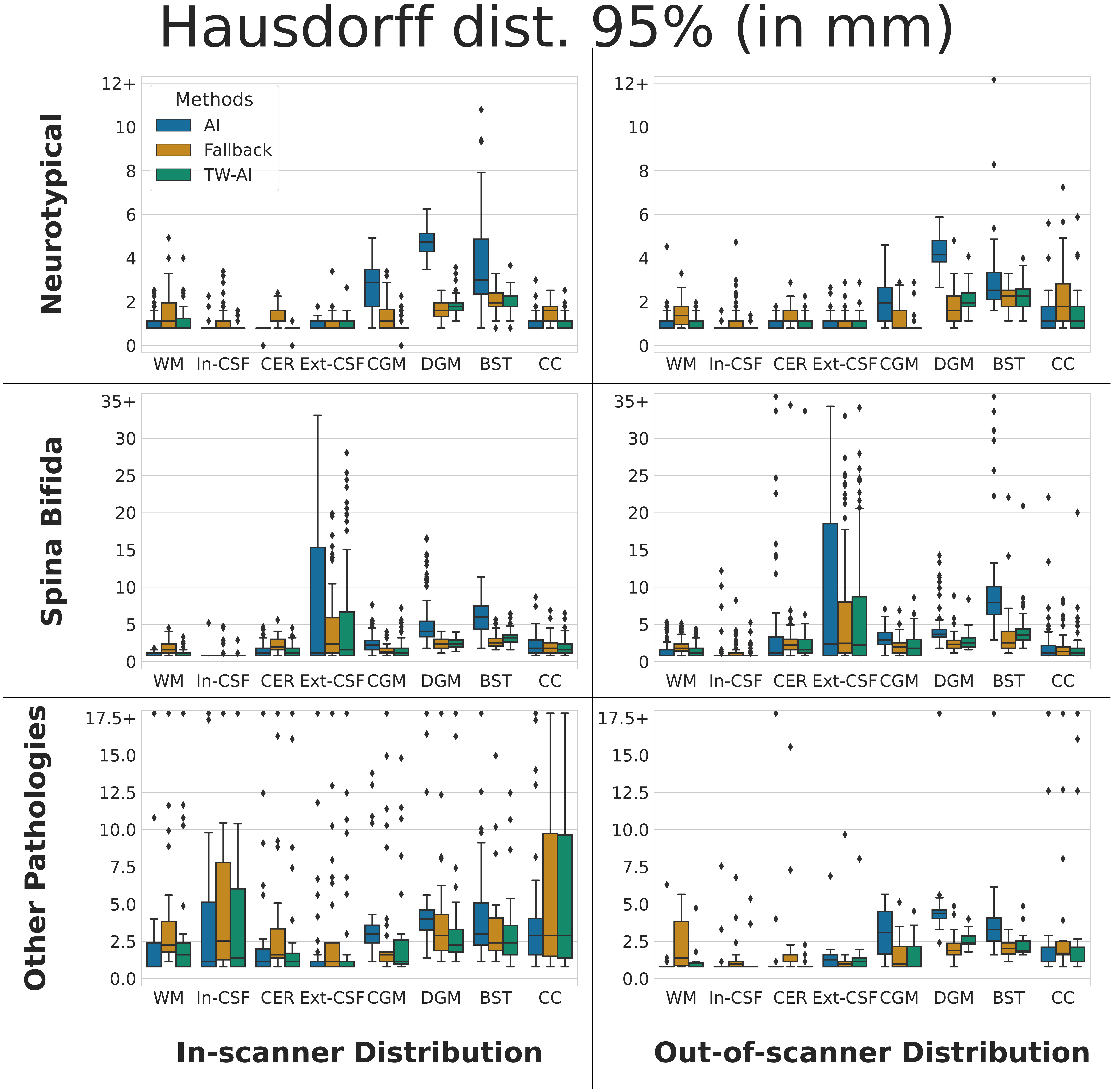}
    \caption{
    \textbf{Hausdorff distance (in mm) comparison of our AI, fallback, and trustworthy AI segmentation algorithms for fetal brain 3D MRI segmentation.}
    The organization and legend of this figure is the same as in Fig.~\ref{fig:dice_roi}, except that here the Hausdorff distance at $95\%$ percentile (HD95) is reported in place of the Dice score.
    To improve the visualization we have clipped the distances to a maximum value. The clipped outliers are still visible on the top of each boxplot.
    \revmod{AI corresponds to nnU-Net~\cite{isensee2021nnu} here.}
    }
    \label{fig:hausdorff_roi}
\end{figure}

\begin{figure}[!htb]
    \centering
    \includegraphics[width=\linewidth,trim=0cm 0cm 0cm 3cm,clip]{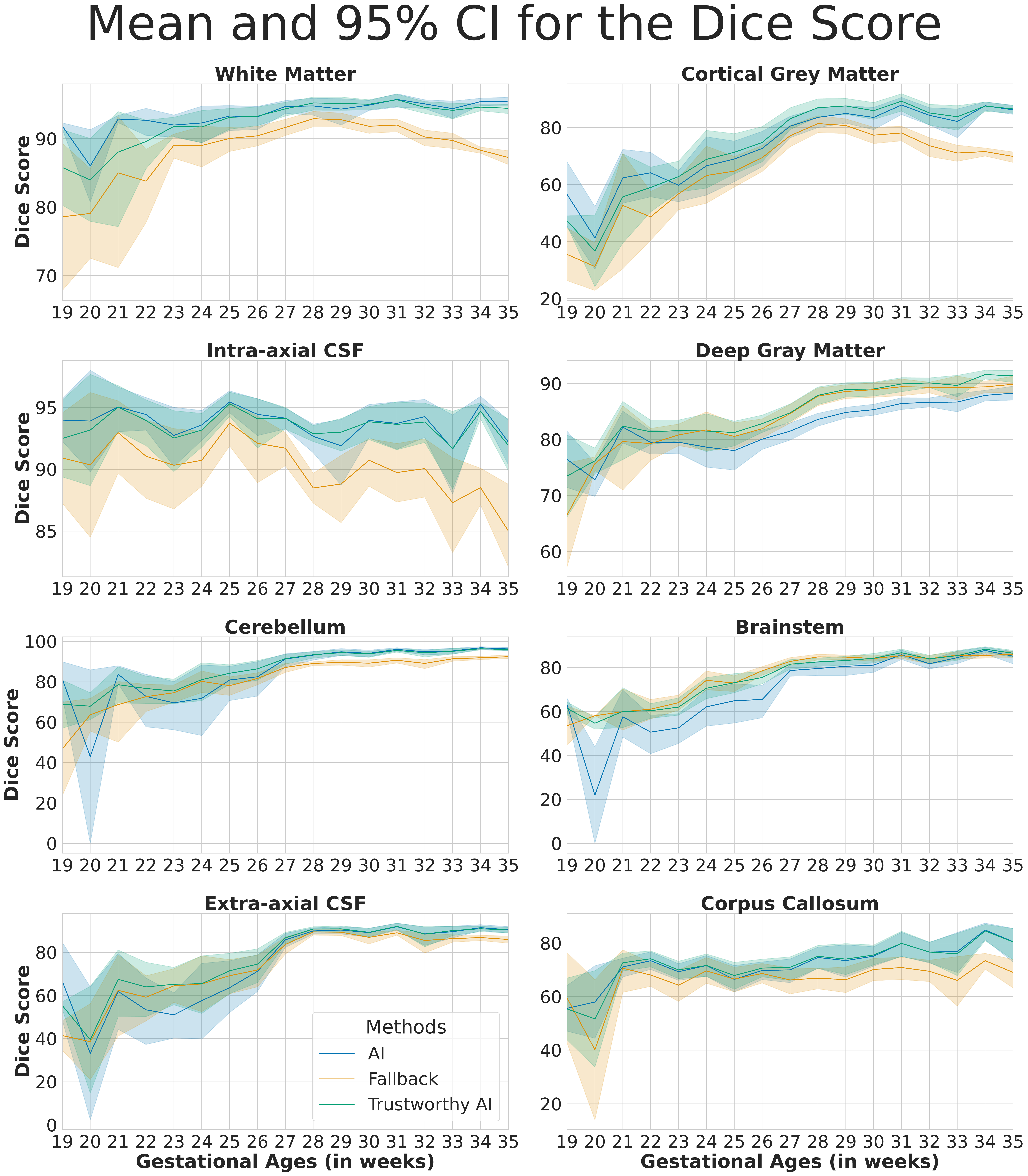}
    \caption{
    \textbf{Mean Dice score (in \%) and $95\%$ confidence interval as a function of the gestational age.}
    Here, we have used all the neurotypical and the spina bifida cases of the testing dataset.
    The trustworthy AI (in green) algorithm achieves similar or higher Dice scores than the best of the \revmod{nnU-Net} backbone AI (in blue) and the fallback (in orange) for all tissue type and all gestational age.
    Fig.~\ref{fig:dice_GA} was obtained from the same data after averaging the scores across regions of interest for each 3D MRI.
    Box limits are the first quartiles and third quartiles. The central ticks are the median values. The whiskers extend the boxes to show the rest of the distribution, except for points that are determined to be outliers.
    Outliers are data points outside the range median $\pm 1.5\times$ interquartile range.
    }
    \label{fig:dice_GA_roi}
\end{figure}

\begin{figure}[!htb]
    \centering
    \includegraphics[width=\linewidth,trim=0cm 0cm 0cm 3cm,clip]{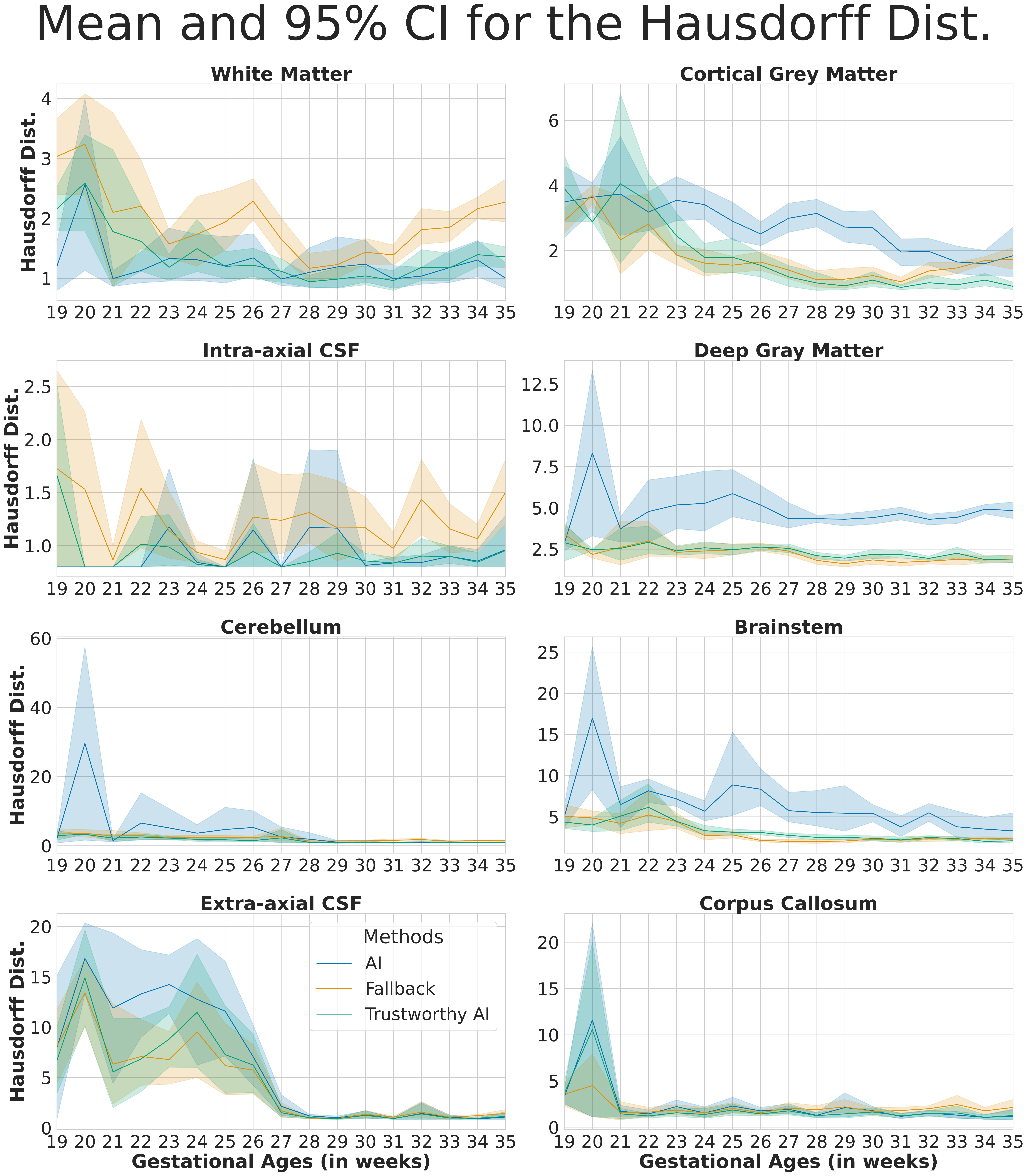}
    \caption{
    \textbf{Mean Hausdorff distance (in mm) and $95\%$ confidence interval as a function of the gestational age.}
    Here, we have used all the neurotypical and the spina bifida cases of the testing dataset.
    The trustworthy AI (in green) algorithm achieves similar or lower Hausdorff distance than the best of the \revmod{nnU-Net} backbone AI (in blue) and the fallback (in orange) for all tissue type and all gestational age.
    Fig.~\ref{fig:hausdorff_GA} was obtained from the same data after averaging the scores across regions of interest for each 3D MRI.
    }
    \label{fig:hausdorff_GA_roi}
\end{figure}

\begin{figure}[!htb]
    \centering
    \includegraphics[width=\linewidth,trim=0cm 0cm 0cm 3cm,clip]{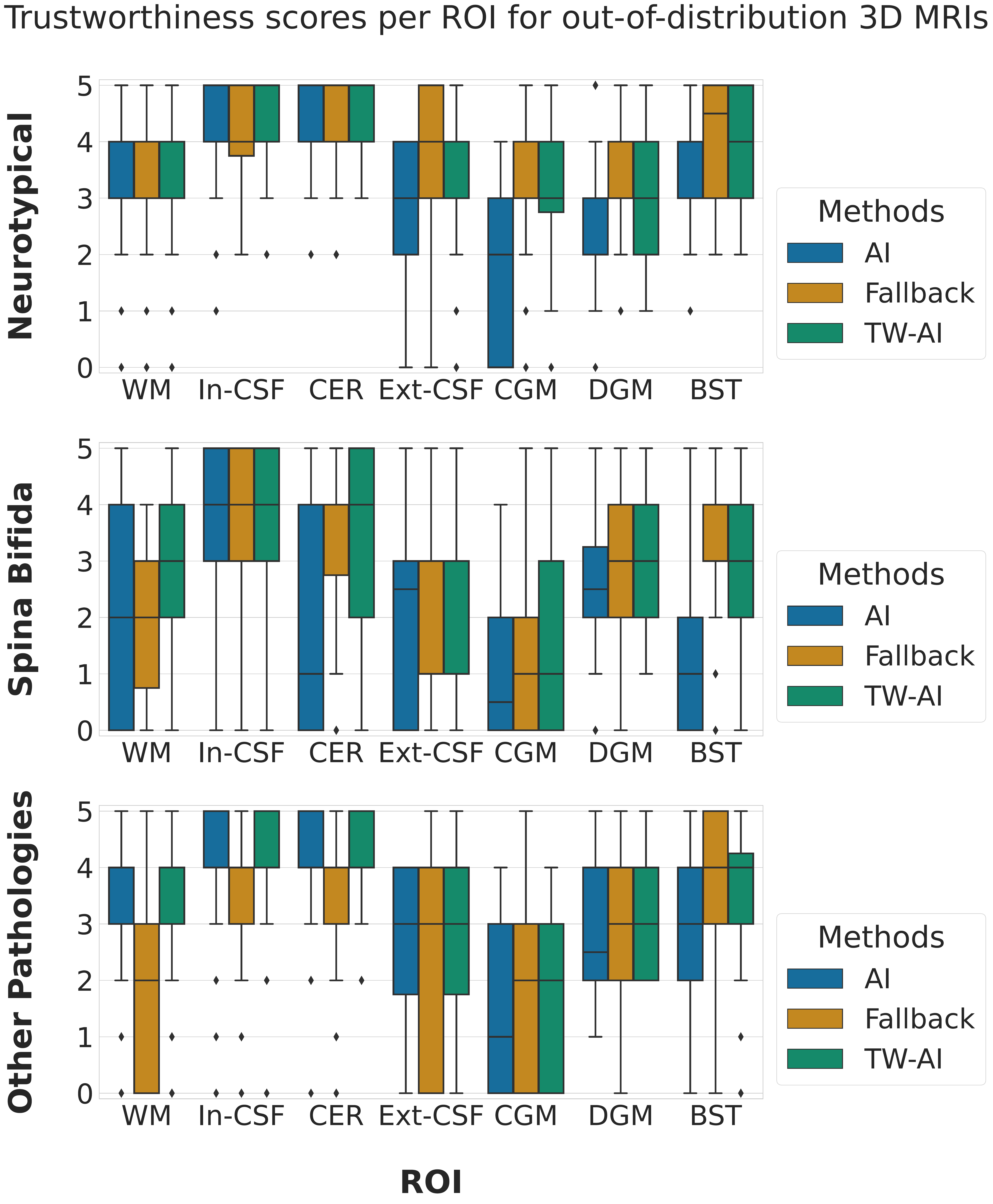}
    \caption{
    \textbf{Experts scores for out-of-distribution 3D MRIs.}
    The scores were evaluated by a panel of \numraters{} experts
    for the three segmentation algorithms for each case and for all the region of interests.
    Experts performed \numscoring{} independent scoring and the algorithms were pseudonymized.
    The scores of different experts for a given region of interest and for a given 3D MRI were aggregated using averaging.
    Here, we have used $50$ out-of-scanner distribution 3D MRIs from the FeTA dataset ($20$ neurotypical, $20$ spina bifida, and $10$ other pathologies).
    %
    %
    \revmod{AI corresponds to nnU-Net here.}
    Fig.~\ref{fig:scores} was obtained from the same data after averaging the scores across regions of interest for each 3D MRI.
    }
    \label{fig:scores_roi}
\end{figure}


\begin{table*}[htb!]
	\caption{\textbf{Tuning of the registration parameters.}
	We report the population average of the mean-class Dice score (DSC) in percentages.
    We also report the average number of volumes that need to be registered for each configuration. This number is approximately proportional to the computational time for the segmentation computation.
    The 3D MRIs used were the fold 0 of the training dataset.
    The row highlighted in \textcolor{ForestGreen}{\textbf{green}} (resp. \textcolor{orange}{\textbf{orange}}) correspond to the value of $\Delta$GA selected for the neurotypical cases (resp. the spina bifida cases).
    A higher value of $\Delta$GA leads to using more volumes in the fallback, registration-based segmentation method.
    Hence, this difference of $\Delta$GA reflects the use of two neurotypical atlases~\cite{gholipour2017normative,wu2021age} while only one spina bifida atlas~\cite{fidon2021atlas} is available.
    }
	\begin{tabularx}{\linewidth}{*{6}{Y}}
	    \toprule
		\textbf{Atlas fusion} & \textbf{Atlas selection} & 
		$\Delta$GA &
		DSC \mbox{Control} & DSC \mbox{Spina bifida}
		& Average \mbox{\#volumes}\\
		\midrule
		Mean & Condition & 0 & 83.9 & 70.0 & 1.6\\
		Mean & Condition & 1 & 84.7 & 72.4 & 4.8\\
		Mean & Condition & 2 & 84.8 & 73.1 & 7.7\\
		Mean & Condition & 3 & 84.9 & 73.3 & 10.4\\
		Mean & Condition & 4 & 85.1 & 73.3 & 13.0\\
		Mean & All       & 0 & 82.0 & 66.4 & 3.1\\
		Mean & All       & 1 & 84.6 & 67.8 & 9.4\\
		Mean & All       & 2 & 85.0 & 66.9 & 15.3\\
		Mean & All       & 3 & 85.1 & 66.8 & 20.8\\
		Mean & All       & 4 & 85.2 & 66.8 & 26.0\\
		GIF  & Condition & 0 & 84.0 & 72.0 & 1.6\\
		\textcolor{ForestGreen}{\bf GIF}  & \textcolor{ForestGreen}{\bf Condition} & 1 & 84.8 & 76.1 & 4.8\\
		GIF  & Condition & 2 & 85.0 & 76.9 & 7.7\\
		\textcolor{orange}{\bf GIF}  & \textcolor{orange}{\bf Condition} & 3 & 85.2 & 77.6 & 10.4\\
		GIF  & Condition & 4 & 85.4 & 77.7 & 13.0\\
		GIF  & All       & 0 & 84.1 & 72.0 & 3.1\\
		GIF  & All       & 1 & 84.9 & 74.6 & 9.4\\
		GIF  & All       & 2 & 85.2 & 75.1 & 15.3\\
		GIF  & All       & 3 & 85.3 & 75.4 & 20.8\\
		GIF  & All       & 4 & 85.5 & 75.3 & 26.0\\
		\bottomrule
	\end{tabularx}
\end{table*}

\subsection{How to implement contracts of trust for image segmentation tasks in general?}
In this section, we discuss two general approaches to implementing a contract of trust for any image segmentation task using either shape prior or intensity prior.

In medical image segmentation of organs and healthy tissues, computed anatomical atlases are typically available as shape priors~\citesupp{iglesias2015multi}.
An image registration method can be used to spatially align each region of interest of the atlas to the image to be segmented.
Due to the spatial smoothness imposed on the spatial transformation, the registered atlas segmentations will
usually be correct up to a spatial margin.
Therefore, a contract of trust can be obtained by adding a spatial margin to the registered segmentation (see Section~\ref{sec:anatomical_contract} and Algorithm~\ref{alg:1} for more details).

Intensity priors present other opportunities to design contracts of trust.
If hyper-intense or hypo-intense voxels are specific to one or several regions of interest, this can be framed as a contract of trust; for example by providing constraints on the composition of a Gaussian mixture model (see Section~\ref{sec:intensity_contract} for more details).

In addition, if more than one contract of trust are available,
the Dempster's rule of combination \eqref{eq:ds} described in Section~\ref{sec:multi_contracts}
provides a simple means of combining them.

\subsection{Toy example: trustworthy traffic lights.}\label{sec:toy-example}
In this section, we provide a toy example to illustrate our method as described in section \ref{sec:multi_contracts}.

One contract of trust for a trustworthy traffic light system at a crossing is that green should not be shown in all directions of the crossing at the same time.
To maintain this contract, traffic light controllers may use a fail-safe conflict monitor unit to detect conflicting signals and switch to a fallback light protocol.
One possible fallback is to display flashing warning signal for all traffic lights.

Formally, for the example of two traffic lights at a single-lane passage, the set of classes is all the pairs of color the two traffic light can display at the same time
$\mathbf{C}=\left\{
(c_1,c_2)\,|\,c_1,c_2 \in \{\textup{green},\textup{orange},\textup{red},\textup{flash}\}
\right\}$.

Let $p^{\textrm{backbone}}$ be the probability of the pair of traffic lights for the default light algorithm.
The probability of the fallback algorithm $p^{\fallback}$ is then defined such as $p^{\fallback}(\textup{flash},\textup{flash})=1$.
The contract of trust is $m$ defined as $m^{\textrm{not-all-green}}\left(\mathbf{C}\setminus \{(\textup{green},\textup{green})\}\right)=1$.

Let $\epsilon \in ]0,1]$, the trustworthy light algorithm is given by
\begin{equation}
    p^{\TWAI}=
    \left((1-\epsilon)p^{\textrm{backbone}}+\epsilon p^{\fallback}\right)
    \oplus m^{\textrm{not-all-green}}
\end{equation}
Using \eqref{eq:drc_proba}, we obtain
\begin{equation}
    \begin{split}
        p^{\TWAI}((\textup{green},\textup{green})) &= 0\\
        p^{\TWAI}((\textup{flash},\textup{flash})) 
        &= \frac{\epsilon p^{\fallback}(\textup{flash},\textup{flash})}{1 - (1-\epsilon)p^{\textrm{backbone}}((\textup{green},\textup{green}))}
    \end{split}
\end{equation}
where the amount of conflict \eqref{eq:contradiction_proba} between $p^{\textrm{backbone}}$ and $m^{\textrm{not-all-green}}$ is equal to $p^{\textrm{backbone}}((\textup{green},\textup{green}))$ and the amount of conflict between $p^{\fallback}$ and $m^{\textrm{not-all-green}}$ is equal to $0$.
In the case of complete contradiction between the default algorithm and the contract of trust, i.e. $p^{\textrm{backbone}}((\textup{green},\textup{green}))=1$, the trustworthy algorithm switch completely to the fallback algorithm with $p^{\TWAI}((\textup{flash},\textup{flash}))=p^{\fallback}(\textup{flash},\textup{flash})=1$.
\subsection{nnU-Net as backbone AI domain generalization segmentation algorithm}\label{sec:nnunet}
The AI segmentation algorithm used is based on nnU-Net~\citesupp{isensee2021nnu} which is a state-of-the-art deep learning-based method for medical image segmentation.
We have chosen the nnU-Net deep learning pipeline because it has lead to state-of-the-art results on several segmentation challenge, including the FeTA challenge 2021 for automatic fetal brain 3D MRI segmentation~\citesupp{fidon2021partial,payette2021automatic}. 
We have used the code available at \url{https://github.com/MIC-DKFZ/nnUNet} without modification
for our backbone AI.

In this section, we give an overview of the nnU-Net deep learning pipeline and of the hyperparameter values selected by nnU-Net for our fetal brain segmentation training dataset.

\textbf{Deep learning pipeline:} The nnU-Net pipeline is based on a set of heuristics to automatically select the deep neural network architecture and other training hyper-parameters such as the patch size.
In this work, a 3D U-Net~\citesupp{cciccek20163d} was selected with one input block, $4$ down-sampling blocks, one bottleneck block, $5$ upsampling blocks, $32$ features in the first level, instance normalization~\citesupp{ulyanov2016instance}, and the leaky-ReLU activation function with slope $0.01$.
This 3D U-Net has a total of $31.2$M trainable parameters.
The patch size selected is $96 \times 112 \times 96$ voxels.

\textbf{Preprocessing:}
We have used the same pre-processing as in our previous work~\citesupp{fidon2021partial}.
The 3D MRIs are skull-stripped using the brain mask after applying a dilation operation ($3$ iterations using a structuring element with a square connectivity equal to one) and setting the values outside the dilated brain mask to $0$.
The brain masks are all computed automatically either during the 3D reconstruction for the Leuven data using \texttt{NiftyMIC}~\citesupp{ebner2020automated,ranzini2021monaifbs}, or for the other data using a multi-atlas segmentation method based on affine registration~\citesupp{modat2014global} and three fetal brain atlases~\citesupp{fidon2021atlas,gholipour2017normative,wu2021age}.
The intensity values inside the dilated brain mask are clipped to the percentile values at $0.5\%$ and $99.5\%$, and after clipping the intensity values inside the dilated brain mask are normalized to zero mean and unit variance.

\textbf{Training:} the training dataset is split at random into $5$ folds.
In total, five 3D U-Nets are trained with one for each possible combination of $4$ folds for training and $1$ fold for validation.
The AI segmentation algorithm consists of the ensemble of those five 3D U-Nets.
Each 3D U-Net is initialized at random using He initialization~\citesupp{he2015delving}.
The loss function consists of the sum of the Dice loss and the cross entropy loss. Stochastic gradient descent with Nesterov momentum is used to minimize the empirical mean loss on the training dataset, with batch size $4$, weight decay $3\times 10^{-5}$, initial learning rate $0.01$, deep supervision on $4$ levels, and polynomial learning rate decay with power $0.9$ for a total of $250{,}000$ training iterations.
The data augmentation methods used are: random cropping of a patch, random zoom, gamma intensity augmentation, multiplicative brightness, random rotations, random mirroring along all axes, contrast augmentation, additive Gaussian noise, Gaussian blurring, and simulation of low resolution.
For more implementation details, we refer the interested reader to~\citesupp{isensee2021nnu} and the nnU-Net \href{https://github.com/MIC-DKFZ/nnUNet}{GitHub page}.

\textbf{Inference:} The probabilistic segmentation prediction of the AI segmentation algorithm is the average of the five probabilistic segmentation prediction of the five 3D U-Nets after training.
In addition, for each 3D U-Net, test-time data augmentation with flip around the $3$ spatial axis is performed.

\revnew{
\subsection{Other backbone AI domain generalization models}\label{sec:other_backbone_AI}
In this section, we give details about the other backbone AI models used in our experiments.

\textbf{SwinUNETR:}
This backbone AI model
is based on a transformer and self-supervised learning pre-training on a large dataset of unlabelled medical images~\cite{tang2022self}.
Self-supervised pre-training has been previously proposed as a domain generalization method~\cite{zhou2022domain}.
SwinUNETR has been reported to outperform nnU-Net on a 3D MRI segmentation task~\cite{tang2022self}.

\textbf{Ensemble nnU-Net + atlas:}
This backbone AI model consists of the average ensembling of nnU-Net and the fallback atlas-based segmentation method.
The same weight equal to $0.5$ is given to each algorithm in the average operation.
%

\textbf{nnU-Net with atlas feature fusion:}
This backbone AI model is based on a fusion method proposed in the domain generalization literature to learn the fusion operation between deep learning features and atlas-based features~\cite{kushibar2018automated,liu2022single}.

We have adapted the nnU-Net deep learning architecture to integrate this fusion method.
Following \cite{liu2022single}, the probabilities of the atlas-based fallback method are concatenated with the deep features of the 3D U-Net at the beginning of the last level of the decoder.
The motivation for performing the concatenation at this layer of the 3D U-Net is to avoid downsampling the atlas-based probabilities which would lead to losing prior information.
At the beginning of the last level of the decoder, the deep features have just been upsampled to the input volume full resolution.

\textbf{Training:}
All backbone AI models were trained using the same training dataset and the same five folds of the training dataset.
nnU-Net with atlas feature fusion follows the exact same training method as nnU-Net as described in section~\ref{sec:nnunet}.
We trained SwinUNETR~\cite{tang2022self} using the code, the pre-trained model, and the instructions of the authors publicly available on their
\href{https://github.com/Project-MONAI/research-contributions/tree/d7bf36c07a0f5882cfddbc7f5aecafea61bf9c39/SwinUNETR/BTCV}{GitHub repository}.
}
\subsection{Multi-atlas segmentation as fallback}\label{sec:fallback}
The fallback segmentation algorithm that we propose to use is based on a multi-atlas segmentation approach.
Multi-atlas segmentation~\citesupp{iglesias2015multi} is one of the most trustworthy approaches for medical image segmentation in terms of anatomical plausibility.
The multi-atlas segmentation that we use is inspired by the Geodesic Information Flows method (GIF)~\citesupp{cardoso2015geodesic}, which is a state-of-the-art multi-atlas segmentation algorithm.

In this section, we give details about the three main steps of the multi-atlas segmentation algorithm used.
First, the selection of the atlas volumes to use to compute the automatic segmentation.
Second, the non-linear registration algorithm to propagate each atlas segmentation to the 3D MRI to be segmented.
And third, the fusion method used to combine the propagated segmentations from the atlas volumes. 

\textbf{Atlas volumes selection:}
We used the volumes from two neurotypical fetal brain 3D MRI atlases~\citesupp{gholipour2017normative,wu2021age} and one spina bifida fetal brain 3D MRI atlas~\citesupp{fidon2021atlas}.
Let GA be the gestational age rounded to the closest number of weeks of the 3D MRI to be segmented. We select all the atlas volumes with a gestation age in the interval $[\textup{GA} - \Delta \textup{GA},\textup{GA} + \Delta \textup{GA}]$ with $\Delta \textup{GA}=1$ week for the neurotypical fetuses and $\Delta \textup{GA}=3$ for spina bifida fetuses.
This way approximately the same number of atlas volumes are used for neurotypical and spina bifida fetuses.

\textbf{Non-linear registration:}
Our image registration step aims at spatially aligning the selected atlas volumes with the 3D MRI to be segmented.
We used \href{https://github.com/KCL-BMEIS/niftyreg}{\texttt{NiftyReg}}~\citesupp{modat2010fast} to compute the non-linear image registrations.
The non-linear image registration optimization problem is the following
\begin{equation}
\label{eq:reg-non-linear}
\left\{
    \begin{aligned}
        &\min_{\Theta} \,\, \mathcal{L}(I_{subject},\, I_{atlas},\, \phi(\Theta)) + R(\Theta)\\
        & R(\Theta) = \alpha_{BE} BE(\phi(\Theta)) + \alpha_{LE} LE(\phi(\Theta))
    \end{aligned}
\right.
\end{equation}
where $I_{atlas}$ is the segmented atlas volume to be registered to the 3D reconstructed MRI $I_{subject}$  that we aim to segment, $\phi(\Theta)$ is a spatial transformation parameterized by cubic B-splines of parameters $\Theta$ with a grid size of $4$ mm.
The data term $\mathcal{L}$ is the local normalized cross correlation (LNCC) with the standard deviation of the Gaussian kernel of the LNCC was set to $6$ mm.
The regularization term $R$ is a linear combination of the bending energy (BE) and the linear energy (LE) regularization functions applied to $\phi(\Theta)$ with $\alpha_{BE}=0.1$ and $\alpha_{LE}=0.3$.

Prior to the non-linear registration, the brain mask of $I_{subject}$ was used to mask the voxels outside the brain and $I_{atlas}$ was registered to $I_{subject}$ using an affine transformation.
The affine transformation was computed using a symmetric block-matching approach~\citesupp{modat2014global} based on image intensities and the brain masks.
The optimization is performed using conjugate gradient descent and a pyramidal approach with $3$ levels~\citesupp{modat2010fast}.
%
%
The hyper-parameters for the non-linear registration were chosen to be the same as in a recent registration pipeline to compute a fetal brain atlas~\citesupp{fidon2021atlas}.

\textbf{Segmentations fusion:}
Once all the atlas volumes $\{I_{k}\}_{k=1}^K$ and their probabilistic segmentations $\{S_{k}\}_{k=1}^K$ have been registered to the 3D MRI to be segmented $I_{subject}$ using the tranformations $\{\phi_{k}\}_{k=1}^K$, we need to fuse the aligned segmentations $\{S_{k}\circ \phi_k\}_{k=1}^K$ into one segmentation.
This fusion is computed via a voxel-wise weighted average using heat kernels~\citesupp{cardoso2015geodesic}.

The heat map for atlas $k$ at voxel $\textbf{x}$ is defined as~\citesupp{cardoso2015geodesic}
\begin{equation}
    w_k(\textbf{x}) = \exp\left(-(D(k, \textbf{x}))^2\right)
\end{equation}
where $D(k, \textbf{x})$ is a surrogate of the morphological similarity between $I_{subject}$ and $I_{k}\circ \phi_k$ at voxel $\textbf{x}$.
The distance $D(k, \textbf{x})$ is the sum of two components
\begin{equation}
    D(k, \textbf{x}) = 
    \alpha L(I_{subject}, I_{k}\circ \phi_k)(\textbf{x})
    + (1 - \alpha) F(\phi_k)(\textbf{x})
\end{equation}
with $\alpha=0.5$, 
$L(I_{subject}, I_{k}\circ \phi_k) = B * \left(I_{subject} - (I_{k}\circ \phi_k)\right)^2$ the local sum of squared differences convoluted (convolution operator $*$) by a B-spline kernel $B$ of order $3$,
and $F(\phi_k)(\textbf{x})$ the Euclidean norm of the displacement field at voxel $\textbf{x}$ (in mm) after removing the low spatial frequencies of $\phi_k$ using a Gaussian kernel with a standard deviation of $20$ mm.
The hyper-parameters chosen are the same as in GIF~\citesupp{cardoso2015geodesic}.
Before computing $D$, the intensity values of the images are normalized to zero mean and unit variance inside the brain mask.

The multi-atlas segmentation at voxel $\textbf{x}$ is computed using the heat kernels as
\begin{equation}
    S_{multi-atlas}(\textbf{x}) = \frac{\sum_{k=1}^K w_k(\textbf{x}) (S_{k}\circ \phi_k)(\textbf{x})}{\sum_{k=1}^K w_k(\textbf{x})}
\end{equation}

Our implementation is available \code{}.

\textbf{Hyper-parameters tuning:}
The hyper-parameters that we tuned are $\Delta\textup{GA}$, the selection strategy for the atlas volumes, and the fusion strategy for combining the probabilistic segmentation of the atlas volumes after non-linear registration.
For $\Delta\textup{GA}$ we tried the values $\{0,1,2,3,4\}$.
For the selection strategy we compared the condition-specific strategy described above to using all the atlases irrespective of the condition of the fetus.
And for the fusion strategy we compared the GIF-like fusion strategy described in above to a simple average.

The data used for the selection of the hyper-parameters are the training data of the first fold that was used for the training of the AI segmentation algorithm.
The mean Dice score across all the segmentation classes and the number of volumes to register were used as our selection criteria to find a trade-off between segmentation accuracy and computational time.
The results can be found in the appendix.
The approach selected consists of GIF-like atlas segmentations fusions, condition specific atlas selection and $\Delta \textup{GA}=1$ for the neurotypical condition and $\Delta \textup{GA}=3$ for the spina bifida condition.

\begin{figure}[t!]
    \centering
    \includegraphics[width=\linewidth]{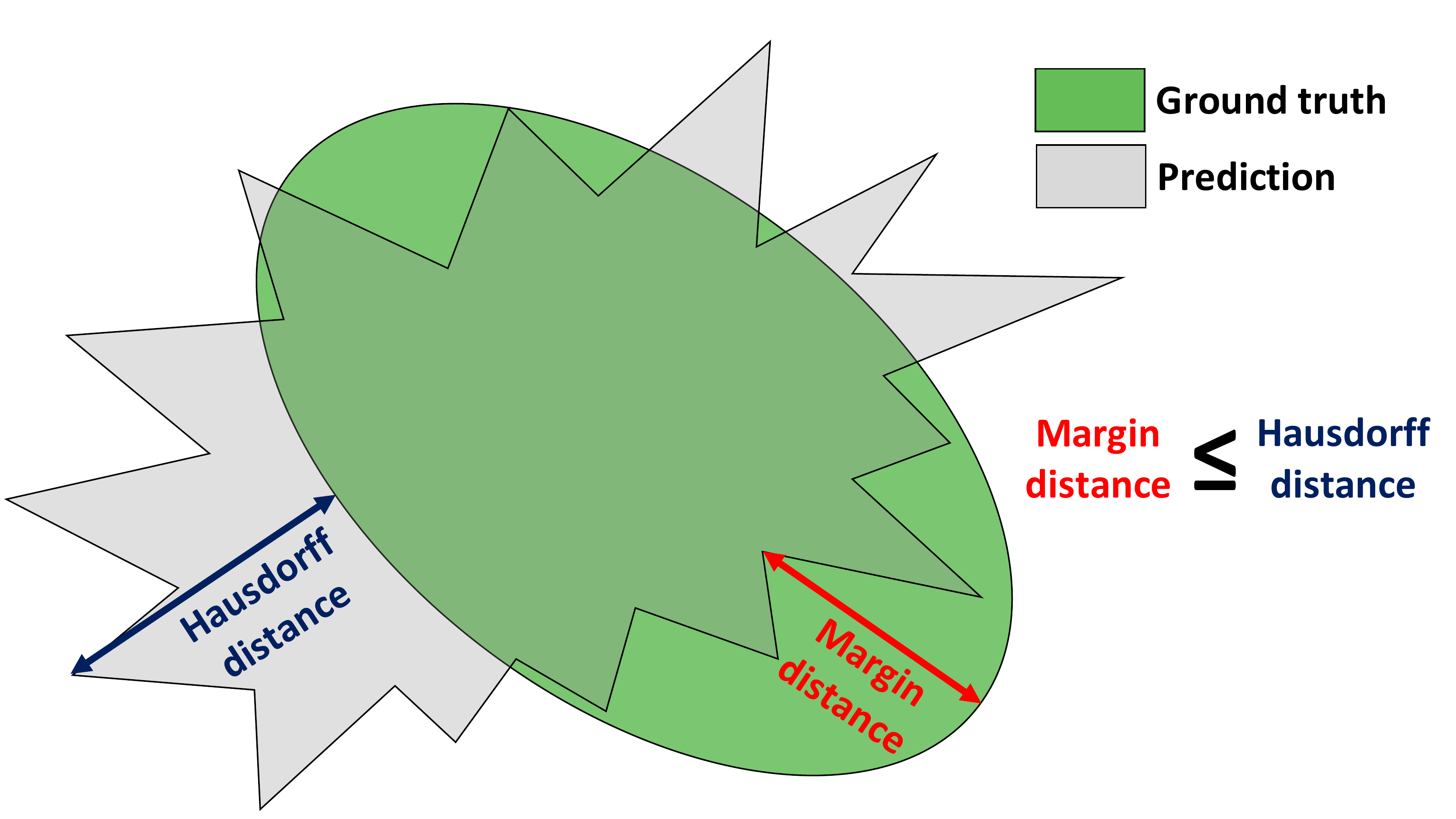}
    \caption{
    \textbf{Illustration of the margin distance.}
    The margin distance is the minimal dilation radius to apply to the predicted binary mask so that it covers entirely the ground-truth binary mask.
    We have proposed to use the margin distance to define our margins used in our definition of the anatomical BPAs.
    }
    \label{fig:margins_definition}
\end{figure}
\revmod{
\subsection{Tuning the margins}\label{sec:tuning_margins}

In our definition of the anatomical prior BPAs \eqref{eq:anatomical_bpa}, the margins account for false negatives in the multi-atlas segmentation.
The anatomical prior BPAs for a class $c$ will impose to the probabilities of class $c$ to be zeros for every voxel outside of the mask after adding the margins using dilation.
Therefore, we chose the margin for a given class $c$ to be the minimal dilation radius for the dilated mask to cover entirely the true region of class $c$ even if it creates overlaps with other regions.

For this purpose, we propose to use a modified Hausdorff distance, called \textit{margin distance}, that considers only the false negatives.
An illustration is given in the appendix, Fig.~\ref{fig:margins_definition}.
Let $\textup{HD}_{95}(M_{pred}, M_{gt})$ denotes the Hausdorff distance at $95\%$ of percentile between a predicted binary mask $M_{pred}$ and the ground-truth mask $M_{gt}$.
The margin distance of interest between $M_{pred}$ and $M_{gt}$ is defined as
\begin{equation}
    \textup{HD}^{FN}_{95}(M_{pred}, M_{gt}) = \textup{HD}_{95}(M_{pred}, M_{pred} \cup M_{gt})
\end{equation}
The margin $\eta_{c,cond}$ for class $c\in\mathbf{C}$ and condition $cond$ (neurotypical or spina bifida) is chosen as the $95\%$ percentile value of $\textup{HD}^{FN}_{95}$ on the fold 0 of the training dataset for the given class and condition.
For fetuses with a condition other than neurotypical or spina bifida, we chose
$\eta_{c,other\,pathologies}= \max\{\eta_{c,neurotypical},\,\eta_{c,spina\,bifida}\}$.
}
\subsection{Implementation details.}
NVIDIA Tesla V100-SXM2 GPUs with 16GB of memory were used for training all the backbone AI deep learning models.
Training each network took from 4 to 6 days for each fold.
Inference of automatic segmentations using the backbone AI models were performed using one NVIDIA GeForce GTX 1070 GPU with 8GB of memory.
On average, the inference time for the backbone AI models was $4$ minutes.

It is worth noting that, for all contracts of trust, the fallback algorithm is used only for inference. 
As a result, our trustworthy AI approach does not have any time or memory overhead during training.

At inference, the fallback algorithm with anatomical and intensity contracts of trust 
ran on CPU using 12 Intel(R) Core(TM) i7-8750H CPU @ 2.20GHz.
On average, this step added $14$ minutes to the inference time. 
The fallback algorithm could benefit from parallelization in two different ways.
First, the segmentation of the fallback could be computed in parallel to the segmentation of the backbone AI model.
Second, the different medical image registrations that dominate the inference time of the fallback algorithm can be parallelized. 
Third, the registration itself could be implemented on the GPU.
These parallelization approaches would bring the inference time of the fallback method well below the inference time of the backbone AI models. 
In this case, the time overhead of our trustworthy AI approach would be reduced to the fusion step which takes only a few seconds (with no particular emphasis having been put on its runtime optimization). 
This extra time is negligible in comparison to the backbone AI inference time.

\subsection{Mathematical notations}
\begin{itemize}
    \item $\mathbf{C}$: the set of all classes to be segmented
    \item $2^{\mathbf{C}}$: the set of all subsets of $\mathbf{C}$
    \item $\textbf{x}$: a voxel or a pixel
    \item $\Omega$: the set of all voxels or pixels (image domain)
    \item $p$: a probability vector
    \item $m$: a basic probability assignment (BPA) in Dempster-Shafer theory
    \item $\oplus$: Dempster's rule of combination
\end{itemize}
\subsection{Proofs of no contradiction}\label{appendix:proof-no-contradiction}

\subsubsection{No contradiction between the anatomical contracts of trust}

Following the assumption of Dempster's rule of combination in \eqref{eq:ds}, we need to make sure that the BPAs $m_c$ defined as in \eqref{eq:anatomical_bpa} are nowhere completely contradictory with each other.

\emph{Proof:}
We show that there is always at least one class that is compatible with the set of anatomical prior.
For this we need to show that for all voxel $\textbf{x}$, there exists $c \in \mathbf{C}$ such that
$
m^{(c)}_{\textbf{x}}\left(\mathbf{C} \setminus \{c\}\right) < 1
\quad
\textup{and}
\quad
m^{(c)}_{\textbf{x}}\left(\mathbf{C}\right) > 0
$.
This holds in our case because the masks $\{M^c\}_{c \in \mathbf{C}}$ form a partition of the set of all the voxels and because, following \eqref{eq:anatomical_bpa}, for the voxels $\textbf{x}$ inside mask $M^c$, $m^{(c)}_{\textbf{x}}(\mathbf{C})=1$ and $m^{(c)}_{\textbf{x}}\left(\mathbf{C} \setminus \{c\}\right)=0$.

\subsubsection{No contradiction between the anatomical prior and the fallback}
For our trustworthy AI model \eqref{eq:trustworhtyAI-fetal} to be valid, we need to show that the probability
$\left((1 - \epsilon) p^{\AI}_{I, \textbf{x}} + \epsilon p^{\fallback}_{I, \textbf{x}}\right)$, 
is not completely contradictory with the anatomical prior BPA.

\emph{Proof:} Since $\epsilon > 0$, it is sufficient to show that $p^{\fallback}_{I, \textbf{x}}$ is not completely contradictory with $m^{\anatomy}_{I, \textbf{x}}$ for every voxel $\textbf{x}$.
In \eqref{eq:anatomical_bpa} and \eqref{eq:anatomical_prior} we have defined the BPA maps $m_{I,c}$ based on the multi-atlas segmentation which is equal to $p^{\fallback}_{I}$.
Therefore, for any voxel $\textbf{x}$, let $c \in \mathbf{C}$ the class such that $\textbf{x} \in M^c$. 
We have $d(\textbf{x}, M^c)=0$ and therefore $m_{I,\textbf{x}}(\mathbf{C})=\phi(d(\textbf{x}, M^c))=1$.
And we have $p^{\fallback}_{I, \textbf{x}}(c)>0$.
This shows that $p^{\fallback}_{I, \textbf{x}}$ and $m^{\anatomy}_{I, \textbf{x}}$ are not completely contradictory.
\subsection{Proof of the formula \texorpdfstring{\eqref{eq:anatomical_BPA}}{(\ref*{eq:anatomical_BPA})} for the anatomical BPA.}\label{appendix:proof_anatomical_BPA}

In this section we give the proof for the formula \eqref{eq:anatomical_BPA}:
For all voxel $\textbf{x}$ and all $\mathbf{C}' \subset \mathbf{C}$,
\begin{equation}
    \label{eq:anatomical_BPA2}
    \begin{aligned}
        &m^{\anatomy}_{I, \textbf{x}}(\mathbf{C}\setminus \mathbf{C}') 
        =\left(\bigoplus_{c \in \mathbf{C}} m^{(c)}_{\textbf{x}}\right)(\mathbf{C}\setminus \mathbf{C}')\\
        &=
        \prod_{c\in \mathbf{C}}
        \left(
        \delta_{c}(\mathbf{C}')m^{(c)}_{\textbf{x}}(\mathbf{C} \setminus \{c\}) 
        + (1 - \delta_{c}(\mathbf{C}'))m^{(c)}_{\textbf{x}}(\mathbf{C})
        \right)
    \end{aligned}
\end{equation}

To simplify the notations and without loss of generality, in this proof we assume that 
$\mathbf{C}=\left\{1,\ldots,K\right\}$ with $K$ the number of classes.
This simply amounts to renaming the classes by the numbers from $1$ to $K$.

Equation \eqref{eq:anatomical_BPA2}, that we want to prove, can then be rewritten as
\begin{equation}
\begin{aligned}
    &m^{\anatomy}_{I, \textbf{x}}(\mathbf{C}\setminus \mathbf{C}') =\\
    &\prod_{c=1}^K
    \left(
        \delta_{c}(\mathbf{C}')m^{(c)}_{\textbf{x}}(\mathbf{C} \setminus \{c\}) 
        + (1 - \delta_{c}(\mathbf{C}'))m^{(c)}_{\textbf{x}}(\mathbf{C})
    \right)
\end{aligned}
\end{equation}

Let us first give the reader an intuition of the formula that we will prove by computing the Dempster's rule of combination for the first two BPAs $m_{\textbf{x}}^{(1)}$ and $m_{\textbf{x}}^{(2)}$.
To simplify the calculations, we will write the Dempster's rule of combination for complements of sets like in \eqref{eq:anatomical_BPA2}.

Let $\mathbf{C}' \subsetneq \mathbf{C}$, using the definition of Dempster's rule of combination \eqref{eq:ds}
and using the relation $\forall \mathbf{G},\mathbf{H} \subset \mathbf{C},\,\, (\mathbf{C}\setminus \mathbf{G})\cap (\mathbf{C}\setminus \mathbf{H}) = \mathbf{C}\setminus (\mathbf{G}\cup \mathbf{H})$
\begin{equation}
    \label{eq:complement_drc}
    \begin{aligned}
    &m_{\textbf{x}}^{(1)} \oplus m_{\textbf{x}}^{(2)}(\mathbf{C}\setminus \mathbf{C}') =\\
    &\quad \frac{\sum_{\mathbf{G},\mathbf{H}\subset \mathbf{C}|\mathbf{G} \cup \mathbf{H} = \mathbf{C}'} m_{\textbf{x}}^{(1)}(\mathbf{C}\setminus \mathbf{G})m_{\textbf{x}}^{(2)}(\mathbf{C}\setminus \mathbf{H})}{1 - \sum_{\mathbf{G},\mathbf{H}\subset \mathbf{C}|\mathbf{G} \cup \mathbf{H} = \mathbf{C}} m_{\textbf{x}}^{(1)}(\mathbf{C}\setminus \mathbf{G})m_{\textbf{x}}^{(2)}(\mathbf{C}\setminus \mathbf{H})}
    \end{aligned}
\end{equation}
Using the definition of $m_{\textbf{x}}^{(1)}$ and $m_{\textbf{x}}^{(2)}$ in \eqref{eq:anatomical_bpa},
$m_{\textbf{x}}^{(1)}(\mathbf{C}\setminus \mathbf{G})=0$ if $\mathbf{G}\not\in \left\{\emptyset, \{1\}\right\}$
and 
$m_{\textbf{x}}^{(2)}(\mathbf{C}\setminus \mathbf{H})=0$ if $\mathbf{H}\not\in \left\{\emptyset, \{2\}\right\}$.

This implies that the sum in the denominator is equal to zeros and that there are only four possible values of $\mathbf{C}'$ such that the numerator is non zeros, i.e. 
$\mathbf{C}'\in \{\emptyset, \{1\}, \{2\}, \{1, 2\}\}$. This gives
\begin{equation}
    \begin{aligned}
        \left(m_{\textbf{x}}^{(1)} \oplus m_{\textbf{x}}^{(2)}\right)(\mathbf{C}) &=
        m_{\textbf{x}}^{(1)}(\mathbf{C}) m_{\textbf{x}}^{(2)}(\mathbf{C})
        \\
        \left(m_{\textbf{x}}^{(1)} \oplus m_{\textbf{x}}^{(2)}\right)(\mathbf{C}\setminus \{1\}) &=
        m_{\textbf{x}}^{(1)}(\mathbf{C}\setminus \{1\}) m_{\textbf{x}}^{(2)}(\mathbf{C})
        \\
        \left(m_{\textbf{x}}^{(1)} \oplus m_{\textbf{x}}^{(2)}\right)(\mathbf{C}\setminus \{1\}) &=
        m_{\textbf{x}}^{(1)}(\mathbf{C}) m_{\textbf{x}}^{(2)}(\mathbf{C}\setminus \{2\})
        \\
        \left(m_{\textbf{x}}^{(1)} \oplus m_{\textbf{x}}^{(2)}\right)(\mathbf{C}\setminus \{1,2\}) &=
        m_{\textbf{x}}^{(1)}(\mathbf{C}\setminus \{1\}) m_{\textbf{x}}^{(2)}(\mathbf{C}\setminus \{2\})
    \end{aligned}
\end{equation}
and $\left(m_{\textbf{x}}^{(1)} \oplus m_{\textbf{x}}^{(2)}\right)(\mathbf{C}\setminus \mathbf{C}')=0$
for all other values of $\mathbf{C}'$.

A general formula is given by, $\mathbf{C}' \subset \mathbf{C}$,
\begin{equation}\tag{$\textup{H}_2$}
    \label{eq:h2}
    \begin{aligned}
        &\left(m_{\textbf{x}}^{(1)} \oplus m_{\textbf{x}}^{(2)}\right)(\mathbf{C}\setminus \mathbf{C}')=
        \\
        &\quad\prod_{c=3}^K\left(1 - \delta_{c}(\mathbf{C}')\right) \times
        \\
        &
        \quad\prod_{c=1}^2
        \left(
        \delta_{c}(\mathbf{C}')m^{(c)}_{\textbf{x}}(\mathbf{C} \setminus \{c\}) 
        + (1 - \delta_{c}(\mathbf{C}'))m^{(c)}_{\textbf{x}}(\mathbf{C})
        \right)
    \end{aligned}
\end{equation}
For clarity we remind that for all $c \in \mathbf{C}$, $\delta_c$ is the Dirac measure associated with $c$ defined as
\begin{equation}
    \forall \mathbf{C}' \subset \mathbf{C}, \quad \delta_c(\mathbf{C}') =
    \left\{
    \begin{array}{cc}
        1 & \textup{if}\,\, c \in \mathbf{C}'\\
        0 & \textup{if}\,\, c \not \in \mathbf{C}'
    \end{array}
    \right.
\end{equation}

The idea of the proof is to generalize formula \eqref{eq:h2} to all the combinations of the first $k$ anatomical BPAs until reaching $k=K$.

For all $k \in \{2,\ldots, K\}$, we defined \eqref{eq:hk} as
\begin{equation}\tag{$\textup{H}_k$}
    \label{eq:hk}
    \begin{aligned}
        &\left(\bigoplus_{c =1}^k m^{(c)}_{\textbf{x}}\right)(\mathbf{C}\setminus \mathbf{C}') =
        \\
        &\prod_{c=k+1}^K\left(1 - \delta_{c}(\mathbf{C}')\right)\times\\
        &\prod_{c=1}^k
        \left(
        \delta_{c}(\mathbf{C}')m^{(c)}_{\textbf{x}}(\mathbf{C} \setminus \{c\}) 
        + (1 - \delta_{c}(\mathbf{C}'))m^{(c)}_{\textbf{x}}(\mathbf{C})
        \right)
    \end{aligned}
\end{equation}
When $k=K$, the set of indices for the first product is empty and the product is equal to $1$ by convention.
Therefore, $\textup{H}_K$ is exactly the same as relation \eqref{eq:anatomical_BPA2} that we want to prove.
We will prove this equality by induction on the variable $k$ for $k$ from $2$ to $K$.

We have already proven \eqref{eq:h2}.
It remains to demonstrate that, for all $k$ from $2$ to $K-1$, $\textup{H}_k$ holds true implies that $\textup{H}_{k+1}$ also holds true.

Let $k \in \{2, \ldots, K-1\}$, let us assume that $\textup{H}_k$ is true.

Let $\mathbf{C}' \subsetneq \mathbf{C}$, using the same formula as in \eqref{eq:complement_drc}
\begin{equation}
    \label{eq:induction1}
    \begin{aligned}
    &\left(\bigoplus_{c =1}^k m^{(c)}_{\textbf{x}}\right) \oplus m_{\textbf{x}}^{(k+1)}(\mathbf{C}\setminus \mathbf{C}') =\\
    &\frac{\sum_{\mathbf{G},\mathbf{H}\subset \mathbf{C}|\mathbf{G} \cup \mathbf{H} = \mathbf{C}'} \left(\bigoplus_{c =1}^k m^{(c)}_{\textbf{x}}\right)(\mathbf{C}\setminus \mathbf{G})m_{\textbf{x}}^{(k+1)}(\mathbf{C}\setminus \mathbf{H})}{1 - \sum_{\mathbf{G},\mathbf{H}\subset \mathbf{C}|\mathbf{G} \cup \mathbf{H} = \mathbf{C}} \left(\bigoplus_{c =1}^k m^{(c)}_{\textbf{x}}\right)(\mathbf{C}\setminus \mathbf{G})m_{\textbf{x}}^{(k+1)}(\mathbf{C}\setminus \mathbf{H})}
    \end{aligned}
\end{equation}

Let us denote
\begin{equation}
    N = \sum_{\mathbf{G},\mathbf{H}\subset \mathbf{C}|\mathbf{G} \cup \mathbf{H} = \mathbf{C}} \left(\bigoplus_{c =1}^k m^{(c)}_{\textbf{x}}\right)(\mathbf{C}\setminus \mathbf{G})m_{\textbf{x}}^{(k+1)}(\mathbf{C}\setminus \mathbf{H})
\end{equation}
Let us first demonstrate that $N=0$.
Using the definition of $m_{\textbf{x}}^{(k+1)}$ in \eqref{eq:anatomical_bpa},
$m_{\textbf{x}}^{(k+1)}(\mathbf{C}\setminus \mathbf{G})=0$ if $\mathbf{G}\not\in \left\{\emptyset, \{k+1\}\right\}$.
Therefore, we need only to study the cases
$G\in \{\mathbf{C}, \mathbf{C}\setminus \{k+1\}\}$.

For $G = \mathbf{C}$, 
$\left(\bigoplus_{c =1}^k m^{(c)}_{\textbf{x}}\right)(\mathbf{C}\setminus \mathbf{G})
= \left(\bigoplus_{c =1}^k m^{(c)}_{\textbf{x}}\right)(\emptyset)
= 0
$ like for every basic probability assignment (BPA).

For $G = \mathbf{C}\setminus \{k+1\}$,
according to \eqref{eq:hk}, that we have assumed true,
\begin{equation}
\label{eq:N0}
    \begin{aligned}
        &\left(\bigoplus_{c =1}^k m^{(c)}_{\textbf{x}}\right)(\mathbf{C}\setminus \mathbf{G})\\
        &\quad= \left(\bigoplus_{c =1}^k m^{(c)}_{\textbf{x}}\right)(\{k+1\})
        \\
        &\quad=\prod_{c=k+1}^K\left(1 - \delta_{c}(\mathbf{C}\setminus \{k+1\})\right)
        \prod_{c=1}^k m^{(c)}_{\textbf{x}}(\mathbf{C} \setminus \{c\})
    \end{aligned}
\end{equation}
We have to consider two cases, $k+1<K$ and $k+1=K$.

If $k+1<K$,
the second term in the first product of \eqref{eq:N0} is equal to $0$
and therfore
$\left(\bigoplus_{c =1}^k m^{(c)}_{\textbf{x}}\right)(\{k+1\})m^{(k+1)}_{\textbf{x}}(\mathbf{C}\setminus \{k+1\})=0$

If $k+1=K$,
we have
\begin{equation}
    \left(\bigoplus_{c =1}^k m^{(c)}_{\textbf{x}}\right)(\{k+1\})m^{(k+1)}_{\textbf{x}}(\mathbf{C}\setminus \{k+1\})
    = \prod_{c=1}^K m^{(c)}_{\textbf{x}}(\mathbf{C} \setminus \{c\})
\end{equation}
Voxel $\textbf{x}$ belongs to at least one of the class binary masks.
Let us denote $c_0$ the binary mask to which is voxel $\textbf{x}$ belongs to.
Using the definition of $m_{\textbf{x}}^{(c_0)}$ in \eqref{eq:anatomical_bpa}, we have 
$m_{\textbf{x}}^{(c_0)}(\mathbf{C}\setminus \{c_0\})=0$.
Therefore, the product above is equal to $0$.
This allows us to conclude, in every case, that $N=0$.

Therefore, \eqref{eq:induction1} becomes
\begin{equation}
    \label{eq:induction2}
    \begin{aligned}
    &\left(\bigoplus_{c =1}^k m^{(c)}_{\textbf{x}}\right) \oplus m_{\textbf{x}}^{(k+1)}(\mathbf{C}\setminus \mathbf{C}') =\\
    &\sum_{\mathbf{G},\mathbf{H}\subset \mathbf{C}|\mathbf{G} \cup \mathbf{H} = \mathbf{C}'} \left(\bigoplus_{c =1}^k m^{(c)}_{\textbf{x}}\right)(\mathbf{C}\setminus \mathbf{G})m_{\textbf{x}}^{(k+1)}(\mathbf{C}\setminus \mathbf{H})
    \end{aligned}
\end{equation}
Similarly as before, due to the definition of $m_{\textbf{x}}^{(k+1)}$, we only need to study the cases of the sets $\textbf{G}$ that are solutions of
$\textbf{G}\cap \emptyset = \textbf{C}'$ or 
$\textbf{G}\cap \{k+1\} = \textbf{C}'$.
The first equality has the unique solution $\textbf{G}=\textbf{C}'$
and the second equality has either no solution, if $k+1\not\in \textbf{C}'$,
or two solutions $G\in\{\textbf{C}'\setminus \{k+1\}, \textbf{C}'\}$ if $k+1\in \textbf{C}'$.
Using the Dirac measure, we can treat all the cases at once and \eqref{eq:induction2} becomes
\begin{equation}
    \label{eq:induction3}
    \begin{aligned}
    &\left(\bigoplus_{c =1}^k m^{(c)}_{\textbf{x}}\right) \oplus m_{\textbf{x}}^{(k+1)}(\mathbf{C}\setminus \mathbf{C}') =\\
    &\left(\bigoplus_{c =1}^k m^{(c)}_{\textbf{x}}\right)(\mathbf{C}\setminus \mathbf{C}')m_{\textbf{x}}^{(k+1)}(\mathbf{C})
    \\
    &+ \delta_{k+1}(\textbf{C}')\left[ \left(\bigoplus_{c =1}^k m^{(c)}_{\textbf{x}}\right)(\mathbf{C}\setminus \mathbf{C}')m_{\textbf{x}}^{(k+1)}(\mathbf{C}\setminus \{k+1\})
    \right.
    \\
    &\left.+ \left(\bigoplus_{c =1}^k m^{(c)}_{\textbf{x}}\right)(\mathbf{C}\setminus (\mathbf{C}'\setminus \{k+1\}))m_{\textbf{x}}^{(k+1)}(\mathbf{C}\setminus \{k+1\})\right]
    \end{aligned}
\end{equation}
Using \eqref{eq:hk}, we can rewrite the second term of \eqref{eq:induction3} as
\begin{equation}
\begin{aligned}
    &\delta_{k+1}(\textbf{C}') \left(\bigoplus_{c =1}^k m^{(c)}_{\textbf{x}}\right)(\mathbf{C}\setminus \mathbf{C}') =
    \\
    &\quad
    \delta_{k+1}(\textbf{C}')
    \prod_{c=k+1}^K\left(1 - \delta_{c}(\mathbf{C}')\right) \times
    \\
    &\quad \prod_{c=1}^k
        \left(
        \delta_{c}(\mathbf{C}')m^{(c)}_{\textbf{x}}(\mathbf{C} \setminus \{c\}) 
        + (1 - \delta_{c}(\mathbf{C}'))m^{(c)}_{\textbf{x}}(\mathbf{C})
        \right)
\end{aligned}
\end{equation}
The product of the first two terms of the product on the right-hand side is
$\delta_{k+1}(\textbf{C}')(1-\delta_{k+1}(\textbf{C}'))=0$, independently to the value of $\textbf{C}'$.
Therefore, the second term of \eqref{eq:induction3} is equal to zeros.

Using \eqref{eq:hk}, and by remarking that 
\begin{equation}
    \begin{aligned}
        \delta_{k+1}(\mathbf{C}'\setminus \{k+1\})
        &=0\\
        \forall c \in \textbf{C}\setminus \{K+1\},\quad \delta_{k+1}(\mathbf{C}'\setminus \{k+1\})
        &=\delta_{k+1}(\mathbf{C}')
    \end{aligned}
\end{equation}
we obtain
\begin{equation}
\begin{aligned}
    &\left(\bigoplus_{c =1}^k m^{(c)}_{\textbf{x}}\right)(\mathbf{C}\setminus (\mathbf{C}'\setminus \{k+1\})) =
    \\
    &\quad
    \prod_{c=k+2}^K\left(1 - \delta_{c}(\mathbf{C}')\right)\times \\
    &\quad\,
        \prod_{c=1}^k
        \left(
        \delta_{c}(\mathbf{C}')m^{(c)}_{\textbf{x}}(\mathbf{C} \setminus \{c\}) 
        + (1 - \delta_{c}(\mathbf{C}'))m^{(c)}_{\textbf{x}}(\mathbf{C})
        \right)
\end{aligned}
\end{equation}

Using this equality and \eqref{eq:hk}, we can rewrite \eqref{eq:induction3} as
\begin{equation}
    \label{eq:induction4}
    \begin{aligned}
    &\left(\bigoplus_{c =1}^k m^{(c)}_{\textbf{x}}\right) \oplus m_{\textbf{x}}^{(k+1)}(\mathbf{C}\setminus \mathbf{C}') = 
    \\
    &
    m_{\textbf{x}}^{(k+1)}(\mathbf{C})
    (1 -\delta_{k+1}(\textbf{C}'))
    \prod_{c=k+2}^K\left(1 - \delta_{c}(\mathbf{C}')\right)\times\\
    &
    \quad \prod_{c=1}^k
        \left(
        \delta_{c}(\mathbf{C}')m^{(c)}_{\textbf{x}}(\mathbf{C} \setminus \{c\}) 
        + (1 - \delta_{c}(\mathbf{C}'))m^{(c)}_{\textbf{x}}(\mathbf{C})
        \right)
    \\
    &+ 
    \delta_{k+1}(\textbf{C}')
    m_{\textbf{x}}^{(k+1)}(\mathbf{C}\setminus \{k+1\})
    \prod_{c=k+2}^K\left(1 - \delta_{c}(\mathbf{C}')\right)\times \\
    &
    \quad \prod_{c=1}^k
        \left(
        \delta_{c}(\mathbf{C}')m^{(c)}_{\textbf{x}}(\mathbf{C} \setminus \{c\}) 
        + (1 - \delta_{c}(\mathbf{C}'))m^{(c)}_{\textbf{x}}(\mathbf{C})
        \right)
    \end{aligned}
\end{equation}
By grouping the two terms we eventually obtain that $\textup{H}_{k+1}$ holds true, i.e.
\begin{equation}
    \begin{aligned}
    &\left(\bigoplus_{c =1}^{k+1} m^{(c)}_{\textbf{x}}\right)(\mathbf{C}\setminus \mathbf{C}') = \\
    &\quad
    \prod_{c=k+2}^K\left(1 - \delta_{c}(\mathbf{C}')\right)\times \\
    &\quad\,
        \prod_{c=1}^{k+1}
        \left(
        \delta_{c}(\mathbf{C}')m^{(c)}_{\textbf{x}}(\mathbf{C} \setminus \{c\}) 
        + (1 - \delta_{c}(\mathbf{C}'))m^{(c)}_{\textbf{x}}(\mathbf{C})
        \right)
    \end{aligned}
\end{equation}

We have proved that $\textup{H}_{2}$ is true and we have proved that for all $k$ from $2$ to $K-1$, $\textup{H}_k$ holds true implies that $\textup{H}_{k+1}$ also holds true.
Therefore, using the induction principle, we conclude that $\textup{H}_{K}$ is true.$\quad \blacksquare$

\subsection{Proof of equality \texorpdfstring{\eqref{eq:DRC_proba_anatomical_BPA}}{(\ref*{eq:DRC_proba_anatomical_BPA})}.}\label{appendix:proof_DRC_proba_anatomical_BPA}

In this section, we give a proof of \eqref{eq:DRC_proba_anatomical_BPA}.
It states that for all $c \in \mathbf{C}$,
\begin{equation}
    \label{eq:DRC_proba_anatomical_BPA2}
    \left(
    p_{I,\textbf{x}} \oplus m^{\anatomy}_{I,\textbf{x}}
    \right)(c) =
    \frac{p_{I,\textbf{x}}(c)m^{(c)}_{\textbf{x}}(\mathbf{C})}{\sum_{c'\in \mathbf{C}}p_{I,\textbf{x}}(c')m^{(c')}_{\textbf{x}}(\mathbf{C})}
\end{equation}

We start the proof from the Dempster's rule of combination for a probability and a BPA \eqref{eq:drc_proba}.
\begin{equation}
\begin{aligned}
    &\left(
        p_{I,\textbf{x}} \oplus m^{\anatomy}_{I,\textbf{x}}
    \right)(c) =\\
    &\quad\quad\frac{p_{I,\textbf{x}}(c)\sum_{\textbf{F}\subset \mathbf{C}|c \in \textbf{F}} m^{\anatomy}_{I,\textbf{x}}(\textbf{F})}{1 - \sum_{c' \in \mathbf{C}} \sum_{\textbf{F} \subset \left(\mathbf{C}\setminus \{c'\}\right)} p_{I,\textbf{x}}(c')m^{\anatomy}_{I,\textbf{x}}(\textbf{F})}
\end{aligned}
\end{equation}
We now rewrite this equation using complement sets in the numerator to be able to use the formula \eqref{eq:anatomical_BPA} for the anatomical BPAs.
\begin{equation}
\label{eq:drc_proof1}
\begin{aligned}
    &\left(
        p_{I,\textbf{x}} \oplus m^{\anatomy}_{I,\textbf{x}}
    \right)(c) =\\
    &\quad\quad\frac{p_{I,\textbf{x}}(c)\sum_{\textbf{G}\subset (\mathbf{C}\setminus \{c\})} m^{\anatomy}_{I,\textbf{x}}(\textbf{C}\setminus\textbf{G})}{1 - \sum_{c' \in \mathbf{C}} \sum_{\textbf{F} \subset \left(\mathbf{C}\setminus \{c'\}\right)} p_{I,\textbf{x}}(c')m^{\anatomy}_{I,\textbf{x}}(\textbf{F})}
\end{aligned}
\end{equation}
Let us first simplify the numerator.
Using \eqref{eq:anatomical_BPA} we obtain, for all $\textbf{G}\subset \mathbf{C}\setminus \{c\}$,
\begin{equation}
\begin{aligned}
    &
    m^{\anatomy}_{I,\textbf{x}}(\textbf{C}\setminus\textbf{G})
    \\
    & 
    =
    \prod_{c'\in \mathbf{C}}
    \left(
        \delta_{c'}(\mathbf{C}')m^{(c)}_{\textbf{x}}(\mathbf{C} \setminus \{c\}) 
        + (1 - \delta_{c'}(\mathbf{C}'))m^{(c')}_{\textbf{x}}(\mathbf{C})
    \right)
    \\
    & 
    =
    m^{(c)}_{\textbf{x}}(\mathbf{C})
    \prod_{c'\in (\mathbf{C}\setminus \{c\})}
    \left(
        \delta_{c'}(\mathbf{C}')m^{(c')}_{\textbf{x}}(\mathbf{C} \setminus \{c'\}) 
    \right.\\
    &\quad\quad\quad\quad\quad\quad\quad\quad\quad\quad \left.
        + (1 - \delta_{c'}(\mathbf{C}'))m^{(c')}_{\textbf{x}}(\mathbf{C})
    \right)
\end{aligned}
\end{equation}
Therefore the term $m^{(c)}_{\textbf{x}}(\mathbf{C})$ can be factorized outside of the sum in the numerator of the right-hand side of \eqref{eq:drc_proof1}.
Let us denote the sum of the numerator, after factorization, as
\begin{equation}
\begin{aligned}
    A_c = \sum_{\textbf{G}\subset (\mathbf{C}\setminus \{c\})}
    \prod_{c'\in (\mathbf{C}\setminus \{c\})}&
    \left(
        \delta_{c'}(\mathbf{C}')m^{(c')}_{\textbf{x}}(\mathbf{C} \setminus \{c'\})
    \right.\\
    &\left.
        + (1 - \delta_{c'}(\mathbf{C}'))m^{(c')}_{\textbf{x}}(\mathbf{C})
    \right)
\end{aligned}
\end{equation}
One can remark that the terms of the product are all independent of $c$.
In addition, the $c'$th terms of the product is either $m^{(c')}_{\textbf{x}}(\mathbf{C} \setminus \{c'\})$ or $m^{(c')}_{\textbf{x}}(\mathbf{C})$ depending on $\textbf{G}$ and when summing over all
$\textbf{G}\subset (\mathbf{C}\setminus \{c\})$ we obtain all the possible products.
Therefore, the sum can be factorized as
\begin{equation}
    A_c =
    \prod_{c'\in (\mathbf{C}\setminus \{c\})}
    \left(
        m^{(c')}_{\textbf{x}}(\mathbf{C} \setminus \{c'\}) 
        + m^{(c')}_{\textbf{x}}(\mathbf{C})
    \right)
\end{equation}
Using the definition of the anatomical BPAs, we have, for all $c \in \textbf{C}$,
$m^{(c')}_{\textbf{x}}(\mathbf{C} \setminus \{c'\}) + m^{(c')}_{\textbf{x}}(\mathbf{C})=1$.
As a result, we obtain $A_c = 1$.

This proves, that 
\begin{equation}
    \left(
        p_{I,\textbf{x}} \oplus m^{\anatomy}_{I,\textbf{x}}
    \right)(c) \propto p_{I,\textbf{x}}(c) m^{(c)}_{\textbf{x}}(\mathbf{C})
\end{equation}
And since
$\sum_{c \in \textbf{C}}\left(p_{I,\textbf{x}} \oplus m^{\anatomy}_{I,\textbf{x}}\right)(c) = 1$,
we can conclude without additional calculations that 
\begin{equation}
    \left(
    p_{I,\textbf{x}} \oplus m^{\anatomy}_{I,\textbf{x}}
    \right)(c) =
    \frac{p_{I,\textbf{x}}(c)m^{(c)}_{\textbf{x}}(\mathbf{C})}{\sum_{c'\in \mathbf{C}}p_{I,\textbf{x}}(c')m^{(c')}_{\textbf{x}}(\mathbf{C})}
\end{equation}
$\blacksquare$

\bibliographystylesupp{IEEEtran}
\bibliographysupp{IEEEabrv,main}

\end{document}